 \documentclass [12pt,a4paper      ]{article}

\usepackage{times}

\DeclareFontFamily{OT1}{times}{}
\DeclareFontShape {OT1}{times}{m }{n }{ <-> ptmr }{}
\DeclareFontShape {OT1}{times}{bx}{n }{ <-> ptmb }{}
\DeclareFontShape {OT1}{times}{m }{it}{ <-> ptmri}{}
\DeclareFontShape {OT1}{times}{bx}{it}{ <-> ptmbi}{}
\usepackage{amsmath}
\usepackage{amsfonts}
\usepackage{amssymb}
\usepackage{latexsym}
\newtheorem{theorem}{Theorem}

\newcommand{\dslash}{\delta  \kern-0.42em \backprime \kern+0.12em} 
\newcommand{\uslash}{\upsilon\kern-0.45em \backprime \kern+0.15em} 
\newcommand*{\seq}{\lim_{n\rightarrow \infty}  \kern-1.10em \backslash \kern+1.10em}  
\newcommand{\cl}{C \kern -0.1em \ell} 
\newcommand{\bbR}{\mathbb{R}}         
\newcommand{\bbC}{\mathbb{C}}         
\newcommand{\bbH}{\mathbb{H}}         
\newcommand{\bbB}{\mathbb{B}}         

\newcommand{\CON}{\overline}          

\newcommand{\Scal}{\mathbb{S}}        
\newcommand{\Vect}{\mathbb{V}}        
\newcommand{\scal}{\circ}             
\newcommand{\vect}{\wedge}            

\newcommand{\SCA}{\langle}            
\newcommand{\LAR}{\rangle}            
\newcommand{\VEC}{\vec{\kern +.1em[}} 
\newcommand{\TOR}{\vec{\kern +.2em]}} 
\newcommand{\BRA}{\langle\kern -.2em\langle} 
\newcommand{\KET}{\rangle\kern -.2em\rangle} 

\newcommand{\Q}{[\hspace{1.mm}]} 
\newcommand{\A}{(\hspace{.5mm})} 

\newcommand{\REV}{\sim}               
\newcommand{\DUA}{\widetilde}         

\newcommand{\ADJ}{\dagger}            
\begin{document}

\title{\bf\vspace{-2.5cm} THE PHYSICAL HERITAGE OF\\ SIR W.R. HAMILTON}

\author{Andre Gsponer and Jean-Pierre Hurni\\ ~\\ 
\emph{Independent Scientific Research Institute}\\ 
\emph{Box 30, CH-1211 Geneva-12, Switzerland.}\\ 
                 ~\\
            ISRI-94-04.27     }

           \date{Presented at the Conference:\\  ~\\
                  {\bf The Mathematical Heritage of \\
                   Sir William Rowan Hamilton}\\ ~\\
              commemorating the sesquicentennial of \\
                 the  invention of quaternions\\ ~\\  ~\\
        Trinity College, Dublin, 17th -- 20th August, 1993}

\maketitle

\begin{abstract}
\noindent
     {\Large 150  years}   after  the  discovery   of   quaternions, 
     Hamilton's    conjecture   that   quaternions   are   a 
     fundamental  language  for  physics is reevaluated  and 
     shown  to be essentially  correct,  provided one admits 
     complex numbers in both  classical and quantum physics, 
     and  accepts  carrying  along the  intricacies  of  the 
     relativistic formalism. Examples are given in classical 
     dynamics,    electrodynamics,   and   quantum   theory.       
     Lanczos's,  Einstein's, and Petiau's generalizations of 
     Dirac's  equation  are  shown  to  be  very   naturally 
     formulated with biquaternions.  The discussion of spin, 
     isospin,  and mass quantization is greatly  facilitated. 
     Compared with other formalisms,  biquaternions have the 
     advantage  of  giving  compact  but at  the  same  time 
     explicit   formulas  which  are  directly  usable   for      
     algebraic or numerical calculations.

\end{abstract}

\newpage

\begin{center}
{\bf \Large Notations}
\end{center}

\noindent In general scalars,  vectors and quaternions are represented  by the following types~:
\begin{itemize}
\item Scalars or biscalars (elements of $\bbR$ or $\bbC$) :\\ Preferably lower case Roman or Greek:  $a, b, c,  \alpha, \beta...$
\item Vectors or bivectors (elements of $\bbR^3$ or $\bbC^3$ ) :\\ Any arrowed character:  $\vec{a}, \vec{B}, \vec{V}, \vec{\omega}, \vec{\pi},...$
\item Quaternions or biquaternions (elements of $\bbH$ or $\bbB$) :\\ Preferably upper case roman or Greek: $A, B, C, \Lambda, \Sigma...$
\end{itemize}

\noindent Square brackets are used to emphasize that bracketed quantities (e.g., $[x]$ for a variable, or $\Q$ for an operand within an expression) are quaternions,  or to represent quaternions as $[scalar; vector]$ pairs :
\begin{itemize}
\item $[x]=[s;\vec{v}]= s+\vec{v} ~~ \in \bbH \text{~~or~~} \bbB$ 
\end{itemize}

\noindent Angle brackets are used to emphasize that bracketed quantities, e.g., $\SCA S \LAR$, are scalars,  or to restrict quaternions to their scalar part :
\begin{itemize}
\item $\SCA s+\vec{v}\LAR = s ~~ \in \bbR \text{~~or~~} \bbC$ 
\end{itemize}

\noindent Please note the use of the following operators :
\begin{itemize}
\item $\CON{\A}$ or $\A^-$  quaternion conjugation (bar or minus)
\item $\A^*$        imaginary  conjugation (star)
\item $\A^+$                 biconjugation (plus)
\item $\A^\REV$ order reversal or ordinal conjugation (tilde) : ~~ $(AB)^\REV = B^\REV A^\REV$
\item $\A^=$  transposition (transpose) : ~~ $\begin{pmatrix} a & b \\ c & d \end{pmatrix}^= = 
\begin{pmatrix} \CON{a} & \CON{c} \\ \CON{b} & \CON{d} \end{pmatrix}$
\item $A \scal B$        scalar part of quaternion product $AB$ : ~~ $A \scal B = \SCA A B \LAR = \Scal[AB]$
\item $A \vect B$        vector part of quaternion product $AB$ : ~~ $A\vect B = \Vect[AB]$
\item $\dot{\A}$       proper time derivative (dot) :  $\dot{\A} = \gamma \dfrac{d}{dict}\A$
\end{itemize}

\noindent With $c$ the velocity of light (we use generally the convention $c=1$) the most common four-dimensional physical quaternions are :
\begin{itemize}
\item     four-position :     $X = [    ict;     \vec{x}  ]$
\item     four-gradient :     $\nabla= 
        [\frac{\partial}{\partial{ict}};\frac{\partial}{\partial\vec{x}}] =
        [\partial_{ict},\vec{\nabla}]$
\item     four-velocity :          $U = \gamma[1;-i\vec{\beta}] $
\item     energy-momentum :             $P = [ E; -i c \vec{p}]$
\item     generalized momentum :        $\Pi = [ H; -i c \vec{\pi}  ]$
\item     electromagnetic four-potential : $A = [\varphi ;    -i \vec{A}]$
\item     charge-current density :      $J = [ \rho ;    -i \vec{j}]$
\item electromagnetic induction bivector  : 
$\vec{F} = [0; \vec{E} +i \vec{B}]$, ~~ $\vec{F}^\REV = [0; \vec{E} -i \vec{B}]$
\item electromagnetic excitation bivector : 
$\vec{G} = [0; \vec{D} +i \vec{H}]$, ~~ $\vec{G}^\REV = [0; \vec{D} -i \vec{H}]$
\item  electromagnetic induction tensor  : $F\A = \A\vec{F} + \vec{F}^\REV\A$
\item  electromagnetic excitation tensor : $G\A = \A\vec{G} + \vec{G}^\REV\A$
\end{itemize}

\noindent Lorentz transformation quaternions operators are `mathcal' characters :
\begin{itemize}
\item Rotation-spinor : $\mathcal{R}\A = \exp(\frac{1}{2}\theta\vec{a})\Q  = [\cos(\frac{1}{2}\theta); \sin(\frac{1}{2}\theta)\vec{a}]\Q$
\item Boost-spinor   : $\mathcal{B}\A = \exp(i\frac{1}{2}y\vec{b})\Q = [\cosh(\frac{1}{2}y); i\sinh(\frac{1}{2}y)\vec{b}]\Q$
\item Lorentz-spinor : $\mathcal{L}\A  =  \mathcal{B}R\Q$
\end{itemize}

\noindent Quaternionic and quantum mechanical scalar products :
\begin{itemize}
\item Angle ``bra'' and ``ket'' delimiters are used to emphasize the range of symbols over which a scalar part is calculated :
$$
\SCA{A|B}\LAR = \SCA{AB}\LAR  = \Scal{[AB]}
$$
\item Formal quantum mechanical Hilbert spaces's scalar products are emphasized by using double ``bra-ket'' symbols, i.e., $\BRA...\KET$ instead of $\SCA ... \LAR$.  E.g.:
 \end{itemize}
$$
 \BRA \psi |...| \psi \KET ~ = ~ \iiint \SCA d^3V ~ \psi^+ (...)\psi \LAR
$$
~~~~~ ~~~ where $\A^+$ is biconjugation, not Hermitian conjugation denoted by $\A^\ADJ$.
\newpage  
                          \begin{center}

           {\bf \large THE PHYSICAL HERITAGE OF SIR W.R. HAMILTON}

               Andre Gsponer and Jean-Pierre Hurni
  
                        {\bf Abstract }

                          \end{center}

     \noindent \emph{150  years   after  the  discovery   of   quaternions, 
     Hamilton's    conjecture   that   quaternions   are   a 
     fundamental  language  for  physics is reevaluated  and 
     shown  to be essentially  correct,  provided one admits 
     complex numbers in both  classical and quantum physics, 
     and  accepts  carrying  along the  intricacies  of  the 
     relativistic formalism. Examples are shown in classical 
     dynamics,     electrodynamics  and   quantum   theory.       
     Lanczos's,  Einstein's, and Petiau's generalizations of 
     Dirac's  equation  are  shown  to  be  very   naturally 
     formulated with biquaternions.  The discussion of spin, 
     isospin,  and mass quantization is greatly  facilitated. 
     Compared with other formalisms,  biquaternions have the 
     advantage  of  giving  compact  but at  the  same  time 
     explicit   formulas  which  are  directly  usable   for      
     algebraic or numerical calculations.} 

                          \begin{center}
                          {\bf Contents}
                          \end{center}
\begin{itemize}
\item[\ref{int:0}.] Introduction.
\item[\ref{def:0}.] Some definitions and properties of quaternion automorphisms.
\item[\ref{irr:0}.] Irreducible representations of biquaternions: Spinors and bispinors.
\item[\ref{rel:0}.] Relativity and quaternionic tensor calculus.
\item[\ref{cla:0}.] Classical dynamics and Hamilton's principle.
\item[\ref{max:0}.] Maxwell's equations and quaternionic analyticity.
\item[\ref{spi:0}.] Spinors in kinematics and classical electrodynamics.
\item[\ref{lan:0}.] Lanczos's generalization of Dirac's equation: Spin and isospin.
\item[\ref{pro:0}.] Proca's equation and the absence of magnetic monopoles. 
\item[\ref{ein:0}.] Einstein-Mayer: Electron-neutrino doublets in 1933!
\item[\ref{pet:0}.] Petiau waves and the mass spectrum of elementary particles.
\item[\ref{qua:0}.] Quaternions and quantum mechanics.
\item[\ref{con:0}.] Conclusion.
\item[\ref{ack:0}.] Acknowledgments.
\item[\ref{bib:0}.] References.
\end{itemize}
\section{Introduction}
\label{int:0}

     For a contemporary physicist, the name Hamilton is primarily 
associated with what is known as the ``Hamiltonian'' formulation of 
dynamics.     Although   mathematically   equivalent   to   other 
formulations such as the Lagrangian formalism,  Hamilton's method 
provides  a  description  of  a classical system  which  has  the 
considerable advantage that the problem can easily be ``quantized,'' 
i.e.,  generalized  from classical to quantum physics.   For this 
reason, in all text books, the operator $H$ appearing on the 
right-hand side of the Schr\"odinger wave equation 
\begin{equation}\label{E1}
  i\hbar \frac{\partial}{\partial t} \Psi  =  H  \Psi,    
\end{equation}
is called the ``Hamilton operator'' or, simply, the ``Hamiltonian.''

     On  its  own,  this formulation of  dynamics  discovered  by 
Hamilton in 1834 (which allows problems of optics and problems of 
mechanics  to  be worked out interchangeably)  is enough  to  put 
Hamilton among the greatest physicists of all times,  at least at 
the level of Newton or Maxwell,  and not very far from  Einstein.       
At this Conference, however, we celebrate another major discovery 
of Hamilton: The invention of quaternions in 1843.

     But quaternions were not Hamilton's only important discovery 
in algebra: Complex numbers were first. Indeed, in 1835, Hamilton 
had  already found a mathematically appealing and consistent  way 
of   interpreting   the  so-called   ``imaginary   numbers.''    By 
considering  pairs of ordinary numbers,  and defining a  suitable 
multiplication rule,  he showed that all operations that could be 
made  with  ordinary numbers could also be made with  his  number 
doublets.

        However,  as  it often happens in the mysterious 
process that we call ``discovery,''  Hamilton had a peculiar mental 
image  in his mind when he was thinking of algebra: 
While   he imagined geometry of a science  of  ``space,''  he 
conceived  algebra  as  the  science  of  ``pure  time''  \cite{R1},  and 
therefore  understanding  imaginary numbers meant for him  coming 
closer to understanding the essence of time.  Thus, when Hamilton 
was  thinking  of  algebra,  his  mental  image  was  that  of  a 
physicist,  an  image of somebody whose ambition is to  discover 
the laws of inanimate nature and motion.

     Therefore,  when after many unsuccessful attempts   Hamilton 
finally succeeded in generalizing complex numbers to quaternions 
(which  require  for their representation not just two  but  four 
ordinary numbers)  he  definitely  believed to have made  a  very 
important discovery.  This conviction,  however,  Hamilton would 
not   include   in  his  scientific   writings.    But   in   his 
correspondence,\footnote{``My letter relates to a certain synthesis of the notions  of Time and Space, ...'' Cited in Ref.~\cite{R1}, page 149.} and in his poetry,\footnote{``And how the One of Time,  of Space the Three,  Might in the 
Chain of Symbol, girdled be.''  Cited in Ref.~\cite{R1}, page 192.} Hamilton made it plain 
that he really thought he had discovered some synthesis of 
three-dimensional space, the vector-part, and time, the scalar-part 
of the quaternion.

    The astonishing fact is that indeed quaternions do foreshadow 
``our four-dimensional world,  in which space and time are  united 
into  a  single  entity,   the  space-time  world  of  Einstein's 
Relativity,'' Ref.~\cite{R2}, page 136.  In effect, as science advances, more 
and  more  evidence  accumulates,  showing that  essentially  all 
fundamental  physics  results can easily  and  comprehensibly  be 
expressed in the language of quaternions. If that is so, then the 
often-made  criticism that Hamilton had ``exaggerated views on the 
importance of quaternions,'' Ref.~\cite{R2}, page 140, was ill-founded.  And 
therefore,  contrary to what is often said,  Hamilton  was right  
to have have spent the last twenty-two years of his life  studying 
all possible aspects of quaternions.

     However,  what  Hamilton  did not know,  and could not  have 
known at his time,  is that quaternions would only become really 
useful in applied and theoretical physics when problems are dealt 
with in which  relativistic  and quantum effects play an essential 
role.   In such problems,  in effect, it is not so much the ``real 
quaternions'' that Hamilton was mostly studying which are useful, but the so-called \emph{biquaternions},  which are obtained when every  four 
components  of  the  quadruplet are  allowed  to  become  complex 
numbers.   This  is  not  to  say that Hamilton's  work  on  real 
quaternions  was  vain.    Quite  the  contrary:  Most  algebraic 
properties of real quaternions that Hamilton so carefully studied 
can be carried over to biquaternions.

     But, for the more practical purposes of non-relativistic 
or  non-quantum  physics  and  engineering,   it  is  true   that 
operations  with quaternions are not sufficiently  flexible.   We 
are therefore  in  a fortunate position  today,  that  after  the 
simplification of the somewhat cumbersome notations used by 
Hamilton,  we  have  at our disposal the modern  vector  notation 
introduced by J.W.\ Gibbs.  Using this notation, we can now work 
with  quaternions much more easily than with Hamilton's  original 
notation.\footnote{When using the modern vector notation with quaternions it is important to keep in mind Gibbs's redefinition of the sign of the scalar product: ~~~ $\vec{a}\cdot\vec{b}  = -\SCA\vec{a}\vec{b}\LAR = -\vec{a} \scal \vec{b} = -\Scal[\vec{a}\vec{b}]$.}   In particular,  either we separate the quaternion into 
its  scalar  and  vector  parts,   separating  or  mixing  freely 
``scalars''  and ``vectors'' \cite{R3},  or use it as a  whole,  especially 
when we deal with the fundamental aspects of theoretical physics, 
as in this paper.

     Hence,  if  we  now anticipate what we will develop  in  the 
sequel,  and jump to our conclusion,  we are indeed going to show 
that  while  quaternions  are  not the panacea  for  solving  all 
possible physical or mathematical problems,  they do nevertheless 
provide an extraordinaryly powerful framework for any problem in 
which  some  four-dimensional or quantal aspect of  our  physical 
world intervenes.

     That this is so, and why it is so, is mysterious.  As  Wigner
stressed  in  his  often  quoted  article  entitled  ``The 
Unreasonable Effectiveness of Mathematics in the Natural Sciences'' 
\cite{R4},  the  biggest  mystery is,  possibly,  the fact that once  a 
particularly efficient mathematical scheme has been found for the 
description of some often crude physical experiment, it turns out 
that the same mathematical tool can be used to give an  amazingly 
accurate description of a large class of phenomena.  This is what 
we  have  discovered:  Once the biquaternion formalism  is  taken 
seriously as a language for expressing fundamental physical laws, 
it  so happens that more and more phenomena  can be predicted  by 
simply generalizing the accepted results while staying within the 
frameworks of quaternions.

     To do so, one has however to follow a few simple guide-lines.  
For instance, the time variable ``$it$''\footnote{Or ``$ict$,''  where $c$ is the velocity of light, to make explicit that the time and space variables have different physical dimensions.} should always be written as 
a pure imaginary number  and,  consequently,  derivation  with 
respect  to  time  should always be  written  $d/dit$.  Hence,  the 
fundamental  space-time  variable $X$ and the corresponding four-gradient $\nabla$  will always be written as
\begin{equation}\label{E2}
        X = [    ict;   \vec{x} ],          
\end{equation}
\begin{equation}\label{E3}
  \nabla = [\frac{\partial}{\partial ict};\frac{\partial}{\partial\vec{x}} ].    
\end{equation}
A  second rule is that there should be no \emph{hidden} ``$i$;''  in other 
words, that the imaginary unit ``$i$'' should always be  explicit, and that imaginary conjugation should always apply to all ``$i$''s.   This 
means  that contrary to the convention of some physicists, e.g., \cite{R3,R5}, one 
should not use the so-called ``Hermitian-'' or  ``Pauli-units,''  but 
only  the  real  quaternion  units  defined  by  Hamilton,  which 
together with the scalar ``$1$'' have the advantage to form a closed 
four-element group, which is not the case with the ``Pauli-units.''

     If these two rules are followed, one discovers that there is 
one  and  only   one  ``$i$'' in physics  and  in  mathematics;  that 
imaginary   conjugation   can  always  be  given   a   consistent 
interpretation  (either  in classical or  quantum  physics);  and
that while ``$i$''\footnote{Or, more precisely, a ``complex structure'' \cite{R6}.} is necessary in quantum theory, ``$i$'' is also a 
very useful symbol in classical physics  because 
it often contributes to distinguish quantities which are of a different 
physical nature.

     To  conclude this introduction,  let us summarize  our  main 
point:   If  quaternions are  used  consistently  in  theoretical 
physics, we get a comprehensive and consistent description of the 
physical  world,  with  relativistic and quantum  effects  easily 
taken  into account.   In other words,  we claim that \emph{Hamilton's 
conjecture},  the  very idea which motivated more then half of his 
professional life, i.e., the concept that somehow quaternions are 
a fundamental building block of the physical universe, appears to 
be  essentially correct in the light of contemporary knowledge.

\section{Some definitions and properties of\\ quaternion automorphisms}
\label{def:0}

     Let  us  review some definitions and properties of the  four 
basic  quaternion  linear  automorphisms,  i.e.,  the  non-trivial 
involutions of $\bbH$,  the field of quaternions,  or $\bbB$ the algebra of biquaternions.

     The first two are \emph{quaternion conjugation}, which reverses the 
sign  of  the  vector  part,  and \emph{imaginary  conjugation},  which 
replaces   the  scalar  and  vector  parts  by  their   imaginary 
conjugate
\begin{equation}\label{E4}
        Q \rightarrow \CON{Q}~  = [ s;  -\vec{v}], 
\end{equation}
\begin{equation}\label{E5}
             Q \rightarrow Q^* = [ s^*; \vec{v^*}].    
\end{equation}
Quite  often  in practice $\CON{\A}$ and $\A^*$ are used in  combination. 
Following  Hamilton's  usage of the prefix \underline{bi}- we  call  this 
third involution \emph{biconjugation}\footnote{Rather than ``Hermitian conjugation,'' symbol $\A^\ADJ$, as it is often improperly called.} 
and use for it the symbol $\A^+$ 
\begin{equation}\label{E6}
        Q \rightarrow Q^+  = [ s^*; -\vec{v^*}] = (\CON{Q})^*.  
\end{equation}
In the same spirit we call a complex vector a  bivector  (rather than a 
``six-vector''),  but  we will refrain from using the term ``biscalar'' 
suggested by Hamilton for complex numbers.

     Using   these  three  involutions  we  have  the   following 
definitions:

\begin{itemize}
\item $Q$ is a scalar if  $\CON{Q}=Q$ ~~~ ~~~ ~~~ ~~~ $\bullet$ $Q$ is a vector   if  $\CON{Q}=-Q$
\item $Q$ is real     if  $Q^*     =Q$ ~~~ ~~~ ~~~ ~~~ ~~~ $\bullet$ $Q$ is imaginary  if  $Q^*     =-Q$
\item $Q$ is bireal   if  $Q^+     =Q$ ~~~ ~~~ ~~~ ~~~ $\bullet$ $Q$ is antibireal if  $Q^+     =-Q$
\end{itemize}

     When  operating  on  a  quaternion  expression,   quaternion 
conjugation reverses the order of the factors.  Thus
\begin{equation}\label{E7}
\CON{AB} = \CON{B} ~ \CON{A}\text{~~~~and~~~~} (AB)^+ = B^+ A^+.  
\end{equation}
     The last non-trivial involution, \emph{order reversal} (or \emph{ordinal conjugation}), 
is more subtle and requires some explanations for which it is best 
to  return  to  Hamilton's ``Elements of  Quaternions,''  and  more 
specifically to a note added by C.J.  Joly in 1898, at the end of 
section ten \cite{R7}, Vol.I, p.162.

     Starting  from  the set of quadruplets of  real  or  complex 
numbers,  the  quaternion algebra is obtained by requiring  their 
product   to  be associative,  and the division to  be   feasible 
always, except possibly in some singular cases. Then, writing two 
quadruplets $A$ and $B$ as scalar-vector doublets $[a;\vec{a}]$ 
and $[b;\vec{b}]$, and using contemporary vector notations, their 
product has the following explicit form
\begin{equation}\label{E8}
   [a; \vec{a}] [b; \vec{b}] = 
   [ ~ \vec{a}\vec{b} + p ~ \vec{a}\cdot\vec{b}~ ;
     ~ a\vec{b}  +  \vec{a}b + q ~ \vec{a}\times\vec{b} ~ ]. 
\end{equation}
The two constants $p$ and $q$ are related by the equation
\begin{equation}\label{E9}
                  q^2  + p^3  = 0,                        
\end{equation}
which  shows  that  there is  some  residual  arbitrariness  when 
defining  the product of two quadruplets.   For instance,  taking 
$p=-1$,  $q$  can  be equal to either $+1$ or $-1$.  On the  other  hand, 
taking  $p=+1$, $q$ may  be $+i$ or $-i$. Thus,  the  choice $p=+1$ 
corresponds to the Pauli algebra.   But, as we have already said, 
we  will keep  Hamilton's  choice, $p=-1$,  which  is  also  more 
fundamental because it corresponds to the Euclidian metric in the 
case of real quaternions,  and to Minkowski's metric in the  case 
where  the four-dimensional space-time position vector is written as 
in formula \eqref{E2}.   Moreover, with Hamilton's choice, the imaginary 
conjugate  of a product is equal to the product of the  imaginary 
conjugate of the factors:  This dispenses of the special rules  which 
are sometimes necessary when using ``Pauli units.''\footnote{Such special rules are also necessary in the standard ``$\gamma$-matrices'' formulation of Dirac's theory because they are based on the Pauli matrices which contain an algebraic $\surd(-1)$  which should not be mistaken with the ``i'' of the complex scalars that multiply the ``$\gamma$''s.}

     In  short,  the arbitrariness in the sign of $p$ is  connected 
with  the signature of the metric,  and the choice of the  metric 
determines  the sign of $p$ and the usage of ``$i$'' when  defining 
physically meaningful four-dimensional quantities.  In effect, since 
the  square of the norm of a quaternion $A$ is by definition  its 
product  by  its  conjugated  quaternion,   we  have   $|A|^2 = A\CON{A}= a^2 + p |\vec{v}|^2$ . Therefore, for a given signature, the choice of the
sign  of $p$ is immaterial because one can  always  multiply  the 
vector part of all quaternions by ``$i$'' in order to get the desired 
signature.

     The  arbitrariness in the sign of $q$ is due to the non-commutativity  
of the quaternion product.   Indeed,  changing the 
order  of the factors $A$ and $B$ is equivalent to changing the  sign 
of $q$.   The involution associated with the changing  of 
this sign is called ``order reversal'' (or simply ``reversal'') and is designated by the symbol $\A^\REV$.\footnote{The operator $\A^\REV$, which is postfixed to an operand, should not be confused with symbol denoting the ``dual'' of an entity, i.e., $\DUA{\A}$.} When biquaternions are used to represent 
physical quantities in space-time, since $q$ is the sign associated 
with  the  vector product,  there is a close  connection  between 
order reversal and space inversion. However, contrary to the 
case of $p$,  there appears to be no invariant overall criterion to 
decide for the sign of $q$.\footnote{This observation relates to the experimental fact that while time flows in only one direction, space is not oriented.}  Therefore,  in accordance  with  the 
principle  of relativity,  one has to make sure that  fundamental 
physical entities are ``order-reversal covariant'' (or simply ``ordinal covariant''), i.e., that they do not arbitrarily depend on the sign of $q$.

     In  this respect a last point is of importance:  Whereas the 
problem  of  signature  is  common  to all formalisms,  order reversal 
is  specific  to  quaternions and Clifford numbers,\footnote{``Order reversal'' is related (but in general not identical) to ``reversion,'' one of the three basic involutions defined on any Clifford algebra.}  and  therefore 
something that should be carefully considered when  any  Clifford 
algebra  is  used  in  physics.  For  this  reason,  because 
reversal was  not  properly  considered  --- for  instance,   in 
defining fundamental quantities as ``ordinal invariant'' --- many 
authors  using  quaternions  in physics have met  with  problems.\footnote{(Note added in 2002.) See, for example, Refs.~\cite{R60,R61}.}  
Indeed,  as  we  will  see,  reversal  plays  an 
essential role when writing fundamental equations of physics.

\section{Irreducible representations of biquaternions:\\ Spinors and bispinors.}
\label{irr:0}

     Before  going to physical applications,  let  us  remind 
(without   proof) a few important  elementary  theorems   concerning   the 
irreducible decompositions of biquaternions:

\begin{theorem}
\label{theo:1}
      Any normed real quaternion, i.e.,  $R=R^*$ such that $|R|^2 =1$,
      can be written 
      \begin{equation}\label{E10}
        R = \exp(\frac{\alpha}{2}\vec{a})  =
        [\cos(\frac{\alpha}{2}); \sin(\frac{\alpha}{2})\vec{a}],   
      \end{equation}
      where $\alpha$ is  a real number called the angle and $\vec{a}$ a  unit 
      vector called the axis.
\end{theorem}
\begin{theorem}
\label{theo:2}
     Any normed bireal quaternion, i.e., $B=B^+$ such that $|B|^2 =1$,
     can be written 
     \begin{equation}\label{E11}
           B = \exp(i\frac{y}{2}\vec{b}) =
               [\cosh(\frac{y}{2}); i\sinh(\frac{y}{2})\vec{b}],   
     \end{equation}
     where $y$ is a real number called the rapidity and $\vec{b}$ a  unit 
     vector called the boost direction.\footnote{A bireal biquaternion is 
     called a \emph{minquat} by Synge \cite{R8}.}
\end{theorem}
\begin{theorem}[Main biquaternion decomposition theorem]
\label{theo:3}
     Any biquaternion $Q$ with non-zero norm, i.e., $|Q|^2 \neq 0$,
     can be written
     \begin{equation}\label{E12}
                          Q = c R B,                        
     \end{equation}
     where $c$ is a complex number, $R$ a normed real quaternion, and  
     $B$ a normed bireal quaternion. 
\end{theorem}
\begin{theorem}
\label{theo:4}
     Any biquaternion $S$ with zero norm,  i.e.,  $|S|^2 = 0$, in
     which  case $S$ is called \emph{singular}\footnote{A singular biquaternion
     is  called a  \emph{nullquat} by Synge \cite{R8}.},  can  be 
     written as a product of three factors
     \begin{equation}\label{E13}
                 Q = r R \sigma,     
     \end{equation}
     where $r$ is a real number, $R$ a normed real quaternion, and 
     $\sigma$ a primitive nullquat 
     \begin{equation}\label{E14}
               \sigma = \tfrac{1}{2} [ 1; i \vec{\nu}] ,      
     \end{equation}
     where $\vec{\nu}$ is a real unit vector, which has the property of 
     being an \emph{idempotent}, i.e., $\sigma^2  = \sigma$.
\end{theorem}
\begin{theorem}
\label{theo:5}
     Multiplying a nullquat from one side by any biquaternion does 
     not change its primitive nullquat.
\end{theorem}

     Since two real parameters are needed to fix the direction of 
a  unit vector,  we see from \eqref{E10} and \eqref{E11} that three  parameters 
are  necessary  to represent a normed real or bireal  quaternion.  
Similarly, a general biquaternion \eqref{E12} requires eight parameters, 
while six suffice for a nullquat \eqref{E13}.

     Theorem \ref{theo:5} provides the basis for defining  spinors in the
quaternion formalism.   In effect,  for a given primitive nullquat 
$\sigma$, the left- (or right-) ideal forms a four parameter group that  is 
isomorph to the spin $\frac{1}{2}$ spinor group \cite{R9}.  As a consequence, 
we have the \emph{spin $\frac{1}{2}$ decomposition theorem} which establishes the link between Dirac's bispinors and biquaternions:

\begin{theorem}[Spin 1/2 decomposition theorem]
\label{theo:6}
     Relative to  a  given primary nullquat $\sigma$, any normed 
     biquaternion Q can be written as a \emph{bispinor}, i.e., as the 
     sum of two conjugated \emph{spinors} 
\begin{equation}\label{E15}
           Q =   R_1 \sigma + R_2 \CON{\sigma},            
\end{equation}
     where $R_1$, $R_2$, are two real quaternions.
\end{theorem}

\section{Relativity and quaternionic tensor calculus}
\label{rel:0}

     Usually,  when special relativity is introduced,  one does a 
lot  of  algebra  in order to work out the  somewhat  complicated 
formulas of Lorentz transformations.   Even when quaternions  are 
used  for  this  purpose,  the proof of the  equivalence  of  the 
quaternion  formulas  with the usual ones is rather  complicated 
\cite{R10}.   In  the present paper,  our intention is to develop   the 
fundamental concepts and present the main results without giving 
the details of the proofs.

     We start therefore from the fundamental ideas of relativity, 
covariance, and tensor calculus which are that all observers are 
equivalent for writing the physical laws, and that all meaningful physical 
quantities  should  have well defined  transformation  properties 
when  going  from  one  observer  to  another  one.\footnote{For an enlightening introduction to tensor calculus and its relation to quaternions see \cite{R11}. See also Ref.~\cite{R2} page 140.}   Hence,  if 
Hamilton's  conjecture is correct,  i.e.,  if indeed biquaternions 
can be used as  elementary building blocs of theoretical physics, 
any meaningful physical quantity should be writable as a  simple 
explicit quaternion expression which should have the same form in all 
{\it referentials}.\footnote{I.e., ``reference frames'' (gallicism).}  In  other words,  the components which  in  ordinary 
tensor calculus are represented by symbols such as $t_{ij...}^{kl...}$, where 
the  various  indices show how the physical quantity varies in  a 
change  of  referential,   should  be  replaced  by  quaternionic 
monomials $QRST...$ where the indices are replaced by some 
convention making the {\it variance}\footnote{I.e., ``transformation law'' (gallicism).} of each factor in the monomial explicit.

     Let   us   take,   for   example,   the   general   Poincar\'e 
transformation  law.  This  is  a change  of  referential   which 
corresponds to the affine function
\begin{equation}\label{E16}
                         Q' = A Q B + C,               
\end{equation}
where  (a  priori)  $A$, $B$ and $C$ are  any  kind  of  quaternion 
expressions. If  $A$ and $B$ are functions of the four-position vector $X$,  
and  the  translation term $C$ is zero,   we have a  local  Lorentz 
transformation,  and  if $A$ and $B$ are independent of $X$  
a  global Lorentz transformation.

     The  most  basic tensor quantity is obviously  the \emph{four-vector}  
such as,  for example,  the space-time vector $X$ given in  formula \eqref{E2}.  Four-vectors have the following variance
\begin{equation}\label{E17}
          V' = \mathcal{L} V \mathcal{L}^+ ,         
\end{equation}
where $\mathcal{L}$ is restricted by Einstein-Minkowski's condition 
\begin{equation}\label{E18}
         \CON{V}V  = \CON{V'}V' = \text{invariant scalar}.   
\end{equation}
Using \eqref{E17} we see that this condition implies
\begin{equation}\label{E19}
\CON{\mathcal{L}} \mathcal{L}= \mathcal{L}^* \mathcal{L}^+ = 1.  
\end{equation}
Thus, the quaternion $\mathcal{L}$ 
which  represents  the most general  Lorentz  transformation,  is 
simply  a  biquaternion  of unit norm.   By theorem  \ref{theo:4},  such  a 
transformation  can therefore be decomposed into a product
    $\mathcal{L} =\mathcal{R}\mathcal{B}$ where $\mathcal{R}$ 
is  a rotation and $\mathcal{B}$ a boost.   To find the explicit form of  the 
boost,   we  apply  \eqref{E17}  to  the  velocity  four-vector  $U$.   Then, 
transforming from the rest-frame (in which $U=1$) to a moving frame, 
we find
\begin{equation}\label{E20}
  U' = \mathcal{B} \mathcal{B}^+ =
       \mathcal{B} \mathcal{B}   = \gamma [1;-i\vec{\beta}], 
\end{equation}
where  $\vec{\beta} = \vec{v}/c$ is the relative velocity of the 
moving frame and 
$\gamma = (1 - \beta^2)^{-1/2}$ 
the Lorentz factor.   We see therefore that  the 
Lorentz  boost is a kind of quaternionic square-root of the 
four-velocity.

     But,  the four-vector is  not  the  most  simple non-trivial 
covariant quantity.  Hence, for a \emph{spinor}, there are four possible 
transformation laws (or eight,  if one takes order reversal into account)
\begin{equation}\label{E21}
   S_1 ' = \mathcal{L} S_1  ~~~~, ~~~~
   S_2 ' = S_2 \CON{\mathcal{L}}  ~~~~, ~~~~
   S_3 ' = \mathcal{L}^*S_3  ~~~~, ~~~~
   S_4 ' = S_4 \mathcal{L}^+. 
\end{equation}
In  fact, as it is immediately seen, these four types of variances  
are the counterparts to the four basic variances of tensor/spinor 
calculus:  Contra- or  co-variance,  dotted or undotted  indices.  
But, here, the last three can be deduced from the first by means of 
the  automorphisms $\CON{\A}$, $\A^*$, and $\A^+$. This leads directly 
to the general idea of quaternionic tensor calculus:   Any time some 
covariant four-dimensional quantity is introduced, the only possible 
new variance it may obtain is the result of operating with one of 
the  three  basic involutions,  possibly  combined  with  reversal
 $\A^\REV$.   Multiplying these quantities, and alternating 
the  variances  by  making use  of  quaternion  conjugation,  one 
obtains more complicated tensors. For example,
\begin{equation}\label{E22}     
                \vec{T} = \Vect[\CON{V_1} V_2],             
\end{equation}
is  a \emph{six-vector}\footnote{Similarly to the concept of \emph{four-vector}, the concept of \emph{six-vector} refers to a variance, not just to the fact that such objects are necessarily bivectors which have six real components.} which has the variance $\vec{T}' = \mathcal{L}^*{\vec{T}}\mathcal{L}^+$. In fact, since 
$\mathcal{L}^* \mathcal{L}^+  = \mathcal{L} \CON{\mathcal{L}} =1$, the scalar part $\SCA \CON{V_1} V_2 \LAR$ of  $\CON{V_1} V_2$ is an invariant, while its 
vector part $\vec{T} =\CON{V_1} \vect V_2$ is a complex vector such as, for example, the electromagnetic field 
bivector $\vec{F} = \vec{E} +i \vec{B} = \CON{\nabla} \vect A$.

     Of  course,  when changing referential,  besides the  proper 
transformations such  as  \eqref{E16},  there  are  also  the  so-called 
improper ones which involve space- or time-reversal, complex conjugation, 
and order reversal.  These can be taken care of in the quaternion 
formalism, using in particular the four basic involutions, so that, 
in the spirit of relativity,  covariance can also be insured with 
respect to them.  For example, in the case of tensors constructed 
by  multiplying  four-vectors of odd parities,  a tensor  that is 
ordinal invariant will also be of odd parity.

     In  summary,  as long as one remains within four-dimensional 
space-time  and  works  with biquaternions,  it  is  possible  to 
achieve   the  same  power  and  flexibility  as  with   ordinary 
tensor/spinor  calculus without having to manipulate explicitly a 
large number of indices.   This is possible in the realms of both 
special and general relativity \cite{R12}.

\section{Classical dynamics and Hamilton's principle}
\label{cla:0}

     Classical  mechanics is a domain in which Hamilton  himself 
found  many  brilliant applications of  real  quaternions.   Just 
think, for example, of his very elegant and general resolution of 
the  Kepler  problem,  in  which the quaternion  formalism  leads 
directly  to  the  conservation of angular momentum  and  of  the 
misnamed ``Runge-Lenz'' vector.\footnote{See Ref.~\cite{R7},  Vol II,  art.  419,  pages 298--299.  The  angular momentum  and Runge-Lenz vectors are denoted by $\beta$ and $\epsilon$, respectively.}

     Here,  however, we will consider the classical dynamics of a 
system  of point particles,  {\it without} specifying a priori whether 
it is a relativistic or non-relativistic problem. The fundamental 
concept is then the ``Hamiltonian,''  which is a scalar function of time, 
position $\vec{x}$, and canonical conjugate momentum $\vec{\pi}$. This 
Hamiltonian 
function  $H$  can be merged together with the  canonical  momentum  
into one bireal quaternion that we call the \emph{four-Hamiltonian}
\begin{equation}\label{E23}
  \Pi = [H(t, \vec{x}, \vec{\pi}); -i \vec{\pi}(t, \vec{x}) ].  
\end{equation}
For  example,  for a system point particles in a time independent 
external field, $\Pi$ is the sum
\begin{equation}\label{E24}
           \Pi  = \sum P_n   +  q_n A(X_n),     
\end{equation} 
where $P_n = [ E_n; -i c \vec{p}_n]$ is the energy-momentum of each particle,  
$q_n$ their electric charge, and $A = [\varphi; -i \vec{A}]$, the external 
electromagnetic four-potential.

     The  equations of dynamics can then be expressed in a number 
of  equivalent forms,  the simplest one being possibly the \emph{action 
postulate} which  states that the four-Hamiltonian derives  from  an 
invariant scalar function, the action $S(t,\vec{x})$ :
\begin{equation}\label{E25}
                        \nabla S = i \Pi.        
\end{equation}
Since $S$ is a scalar,  operating on both sides with  $\CON{\nabla}$ 
and taking the vector part we find
\begin{equation}\label{E26}
         \CON{\nabla} \wedge \Pi  =  0.   
\end{equation}
This  is,  written in quaternions,  the condition for $d\Pi$ to be  a 
total differential, or, equivalently, for $\Pi$ to be an exact one-form, i.e.,
\begin{equation}\label{E27}
        \oint  \CON{dX} \scal \Pi   =  0,     
\end{equation}
which,  for any given two fixed point $X_1$ and $X_2$, 
is the same as \emph{Hamilton's principle}
\begin{equation}\label{E28}
    \delta  \int_{X_1}^{X_2}  \CON{dX} \scal \Pi   =  0.   
\end{equation}
Moreover,  for a system in which $H$ does not depend explicitly on 
time,  \eqref{E27} is also fully equivalent to \emph{Hamilton's equations} of 
motion:
\begin{equation}\label{E29}
   \dot{\vec{x}} = \frac{d}{d\vec{\pi}} H ~~~, ~~~~~   \dot{\vec{\pi}} = -\frac{d}{d\vec{x}} H.  
\end{equation}
The  sequence  of transformations we have gone through  may  look 
like a succession of trivialities.   This,  in fact,  is not  the 
case. Had we not put the ``$i$'' in front of $\vec{\pi}$ in \eqref{E23}, we would  not  
have been able to work out  these results.   Moreover,  while all 
expressions  are formally covariant,  they are the same whether we 
assume  the  kinematical expression for $P$ to be  relativistic  or 
not.  Hence,  although  we have done nothing more than  rewriting 
well  known results,  we see that complex quaternions  provide  a 
compact  and  convenient framework for writing the  equations  of 
Hamiltonian  dynamics,  and  that the resulting  expressions  are 
automatically  in relativistic covariant form.   It is the same with 
Maxwell's equations:  As will be recalled in the next section, writing them 
down in compact quaternionic form requires the use \underline{bi}quaternions.

     To show what happens if we introduce relativity suppose  for 
example that  we apply \eqref{E25} to  a  single  particle.  The  fourth 
component,  i.e.,  time  or  energy,  is no longer an  independent 
variable.  For  instance,  we have now the relativistic  identity 
$|P|^2 = m^2$  and \eqref{E25} can be rewritten as
\begin{equation}\label{E30}
   (i \CON{\nabla} S + e \CON{A}) (i \nabla S + e A) =  m^2.   
\end{equation}
This is \emph{Hamilton-Jacobi}'s  equation for a  particle in an 
electromagnetic field.

\section{Maxwell's equations and quaternionic analyticity}
\label{max:0}

After  the  casting  of Lorentz transformations  into  quaternion 
form,  one of the first modern applications of biquaternions  was 
the rewriting of Maxwell's equations in 1911 by Conway \cite{R13}, and in 1912 by Silberstein \cite{R14}, as
\begin{equation}\label{E31}
\left.
\begin{array}{c}
\CON{\nabla} A = \vec{F}, \\    
\nabla \vec{F} = - 4 \pi  J,
\end{array}
\quad \right\}
\end{equation}
where  $\vec{F} = \vec{E}+i\vec{B}$ is the electromagnetic field {bivector}, $J = [ \rho, -i\vec{j}]$ the source charge-current density,
and $A = [\varphi, -i\vec{A}]$ the electromagnetic potential for which
the Lorenz gauge, i.e., $\CON{\nabla} \scal A = 0$, is assumed.  
This very compact form  allows many 
calculations  to  be done very effectively.   In  particular,  as 
shown  by Silberstein in 1913, the energy-momentum tensor  of 
the electromagnetic field,  i.e.,  Maxwell's stress-energy tensor, 
has a very simple explicit quaternionic form \cite{R15}
\begin{equation}\label{E32}
  4 \pi  T\A =  \tfrac{1}{2} \vec{F}^+    \Q \vec{F}
             = -\tfrac{1}{2} \vec{F}^\REV \Q \vec{F},  
\end{equation}
where  the  free space $\Q$ corresponds to the  position  of  a quaternionic 
argument.\footnote{This convention, due to Hamilton and promoted by Silberstein, Conway and Synge \cite{R8}, generalizes Dirac's ``bra--ket'' notation to biquaternions.  Its value stems from the speed of calculation which derives from the simplicity of the composition rule: $a \Q a' \odot b \Q b' = ab \Q b'a'$~. Moreover, it provides a clear distinction between ``numbers'' (or ``vectors'') $Q$, and ``functions'' (or ``operators'')  $Q\Q$~.} When 
this  tensor is used to calculate  the  flow  of 
energy  and momentum through an hypersurface  the result 
is automatically covariant, and there is no ``4/3 problem'' as  with 
the  obnoxious  Poynting vector \cite{R16}.   Moreover,  using \eqref{E31}  to 
calculate  its  divergence, one immediately obtains Lorentz's force-density 
equation
\begin{equation}\label{E33}
\frac{d \dot{P}}{d^3V} = - T(\nabla)= -\tfrac{1}{2} (J\vec{F} + \vec{F}^\REV J)
                       = -F(J),  
\end{equation}
which shows that the quaternion form of the electromagnetic field 
\emph{tensor} is $F\A = \tfrac{1}{2} (\Q \vec{F} + \vec{F}^\REV\Q)$, a physical object which should not be confused with the electromagnetic field \emph{bivector} $\vec{F} = \vec{E}+i\vec{B}$, or its reverse  $\vec{F}^\REV = \vec{E}-i\vec{B}$, a non-trivial distinction first made in 1955 by Kilmister \cite{R17}.

     A  most interesting idea suggested by Maxwell's equations in 
quaternion  form was developed by Lanczos \cite{R18} in his PhD  thesis 
of 1919.   In effect,  Maxwell's second equation in vacuum,  $\nabla \vec{F} = 0$, is  
the  direct  generalization  of  the  Cauchy-Riemann  analyticity 
condition from two to four dimensions.  It is therefore natural to 
envisage  classical  electrodynamics as  a  biquaternionic  field 
theory  in which point singularities are interpreted as electrons 
\cite{R19}.   In  this case the field at some point $X$ is calculated  by 
means  of  the appropriate generalization  of  Cauchy's  formula 
in which  the  integration contour becomes an  hypersurface
 $\Sigma(Y)$ surrounding the point
\begin{equation}\label{E34}
 \vec{F}(X)  = \frac{-1}{2\pi^2} \iiint \frac{\CON{R}}{|R|^4}
                                  ~ d^3{\Sigma} ~ \vec{F}(Y),   
\end{equation}
where $R=Y-X$ and $|R|^2 = \CON{R}R$. This generalization of complex analysis 
has  been  extensively  studied  by Fueter in the  case  of  real 
quaternions \cite{R20} and more recently extended to biquaternions  and 
higher  dimensional Clifford algebras \cite{R21}.   This formalism  can 
now very efficiently be applied to standard problems, such as the 
calculation of retarded potential and fields \cite{R22}.

\section{Spinors in kinematics and\\ classical electrodynamics}
\label{spi:0}

     Spinors are increasingly often used in classical physics and 
relativity \cite{R23}.   However, possibly the first significant use of 
spinors in classical physics was made in 1941 by Paul Weiss \cite{R24}, the 
particularly brilliant PhD student of Max Born.  By ``significant'' 
we  mean that Weiss's applications of quaternions was not just  
rewriting an otherwise known result in quaternion form.   In fact, 
Weiss gave an independent interpretation  and  derivation  of  an 
important physical law:  The Lorentz-Dirac equation.

     Weiss's starting  point  was the fact  that  the  quaternion 
formalism  provides  explicit  formulas which  are  difficult  to 
obtain  by the ordinary methods of analysis.   For  instance,  in 
kinematics,  taking the square root of the four-velocity as in  \eqref{E20} 
is the same as making the spinor decomposition of the four-velocity. 
An  explicit  formula for the four-acceleration is then obtained  by 
taking the total proper-time derivative on both sides of
\begin{equation}\label{E35A} 
         \dot{Z} = U = \mathcal{B} \mathcal{B}^+,     
\end{equation}
which leads to
\begin{equation}\label{E35B} 
         \ddot{Z} = \dot{U} = - \mathcal{B} \vec{\mathcal{A}} \mathcal{B}^+,     
\end{equation}
where, as shown by Weiss, the invariant real vector $\vec{\mathcal{A}}$
 is the acceleration in the instantaneous  rest-frame.


   Similarly,  since null-four-vectors can explicitly  be formulated with biquaternions,  
one has explicit formulas for the light-cone and retarded coordinates, i.e,
\begin{equation}\label{E36}
     X-Z = 2i \xi \mathcal{B} \CON{\sigma} \mathcal{B}^+,       
\end{equation}
where $\xi=-iU\scal\CON{(X-Z)}$ is the invariant retarded distance from the position of the charge $Z$ to the space-time point $X$, and $\sigma$ an idempotent such as \eqref{E14} with $\vec{\nu}$ pointing from $Z$~to~$X$.

      In  his  paper,  Weiss does not speak of  spinors.  On  the 
contrary, he makes it clear that his decomposition has nothing to 
do  with Dirac's \emph{bispinors}.  But what he does is exactly the kind 
of spinor decomposition we use today, e.g., in general relativity.

     Weiss's application is to show that in this formalism the flow 
of energy and momentum through a hypersurface surrounding a point 
charge  in  arbitrary  motion  can be  calculated  exactly  using 
Silberstein's form of the Maxwell tensor \eqref{E32}.   He then proceeds 
to  find  the world-lines for which the energy-momentum  flow  is 
stationary,  and discovers that the resulting equation of  motion 
is nothing but the Lorentz-Dirac equation \cite{R25}
\begin{equation}\label{E37}
 mc^2 \dot{U} = \tfrac{2}{3} i e^2(\dot{U} \dot{\CON{U}} U + \ddot{U}) -\tfrac{1}{2} e(U \vec{F} + \vec{F}^\REV U). 
\end{equation}

\section{Lanczos's generalization of Dirac's equation:\\ Spin and isospin.}
\label{lan:0}

     About  one  year  after  Dirac discovered  his  relativistic 
wave-equation for spin $\tfrac{1}{2}$ particles,  Lanczos \cite{R26} published a 
series of three articles in which he showed how to derive Dirac's 
equation from the more fundamental coupled biquaternion system\footnote{We write $m$ for $mc/ \hbar$  taking  $c=\hbar=1$ for  simplicity. Note that Lanczos could have taken the reverse of \eqref{E38} as his fundamental equation:$~~~ ~~~ ~~~ A \CON{\nabla} = m B^\REV  ~~~ , ~~~ B^\REV\nabla = m A. ~~~ ~~~ ~~~ (\ref{E38}^\REV)$ }
\begin{equation}\label{E38}
\left.
\begin{array}{c}
\CON{\nabla} A = m B, \\    
\nabla B = m A.
\end{array}
\quad \right\}
\end{equation}
Obviously,  Lanczos  was  inspired  by  his  previous  work  with 
quaternions \cite{R18}.  Indeed,  comparing with \eqref{E31}, it is clear that \eqref{E38} 
can  be  seen  as Maxwell's equations with  feedback,  and  that, 
following  Lanczos \cite{R27},  this feedback can be interpreted  as  a 
distinctive  feature  of massive  particles.  However,  there  is 
a problem. In Dirac's equation, the wave function is a four-component 
bispinor,  while  $A$ and $B$ are biquaternions with  four 
complex  components each.   This is the ``doubling''  problem  that 
puzzled  Lanczos  a lot,  as well as  others who later tried  to 
cast Dirac's equation in quaternion form \cite{R28}.

     The first step towards a Dirac bispinor is to postulate that 
$A$ and $B$ have spinor variances,  i.e., that  $A'= \mathcal{L}A$ and $B'= \mathcal{L}^* B$,  which 
leaves the possibility of making a gauge transformation,  i.e.,  a 
right-multiplication by some arbitrary biquaternion $G$.   Then, to 
get  Dirac's spin $\tfrac{1}{2}$ field, i.e., a bispinor, Lanczos had to make the superposition (see theorem \ref{theo:6} in Sec.~\ref{irr:0})
\begin{equation}\label{E39}
                D =  A \sigma  +  B^* \CON{\sigma}.  
\end{equation}
Here $\sigma$ is an idempotent such as \eqref{E14}, where  
$ \vec{\nu}$ is any unit quaternion.  Comparing with \eqref{E15},  we see 
that $\sigma$ has the effect of projecting out half of $A$, which added to 
another half of the complex conjugated of $B$,  gives  a  Lorentz 
covariant superposition that obeys the wave equation
\begin{equation}\label{E40}
             \CON{\nabla} D = m D^* i \vec{\nu}.   
\end{equation}
This equation, to be called  the \emph{Dirac-Lanczos  equation},  is 
precisely   equivalent   to  Dirac's  equation.    It   will   be 
rediscovered  by  many people,  in particular by G\"ursey \cite{R29} and 
Hestenes \cite{R30}.   While equivalent to other possible  forms,  \eqref{E40} 
has the considerable didactic advantage of making ``spin'' explicit.  
Indeed,  the unit vector on the right-hand side shows  that  Dirac's 
equation  singles  out  an  arbitrary  but  unique  direction  in 
ordinary space:  The \emph{spin} quantization axis.\footnote{(Note added in 2006.) Even more important, the complex conjugation on the right-hand side explicitly shows that the Dirac field is fermionic, rather than bosonic as the Maxwell and Proca fields \cite{R27,R61}.}

     Using this equation,  it is easy to construct and study  the 
various  covariant  quantities  which are  important  in  quantum 
electrodynamics.   For example, the conserved probability current 
is $J=DD^+$, and Tetrode's energy-moment tensor is
\begin{equation}\label{E41} 
                T\A = \bigl(\SCA \Q \CON{\nabla}\LAR D\bigr) i\vec{\nu} D^+ -
 Di\vec{\nu} \bigl(D^+ \SCA\CON{\nabla}\Q \LAR\bigr) -
                      \SCA\CON{A}_{\text{em}}\Q \LAR DD^+,        
\end{equation}
where $A_{\text{em}}$ is the potential of an external electromagnetic field.

     However,  the superposition \eqref{E39} is not the only one leading 
to  a  spin $\tfrac{1}{2}$ field obeying equation \eqref{E40}.   As shown by  G\"ursey  in 1957,  if \eqref{E39} represents a \emph{proton},  the \emph{neutron} is then \cite{R31,R32}
\begin{equation}\label{E42}
               N = ( A \CON{\sigma} - B^* \sigma) i \vec{\tau},     
\end{equation}
where $\vec{\tau}$ is any unit vector orthogonal to $\vec{\nu}$.

     Hence,  Lanczos's doubling is nothing but \emph{isospin}.   G\"ursey's 
articles  had  a tremendous impact \cite{R27} and inspired  ideas  like 
chiral  symmetry  and the sigma model  \cite{R32}.  Indeed,  ``internal'' 
symmetries  such as isospin are explicit and trivial in  Lanczos's 
double  equation  \eqref{E38},   while  only  space-time  symmetries are 
explicit  in  Dirac's traditional $4 \times 4$ matrix formulation  or  the 
biquaternionic formulation \eqref{E40}.   Unfortunately,  except in  his 
PhD dissertation,  G\"ursey made no reference to Lanczos's work, and 
Lanczos  never  learned  that he had anticipated isospin  in  1929 
already!

     Since the fundamental fields are $A$ and $B$, while $D$ and $N$ are 
the physically observed ones,  it is of interest to find the most 
general  gauge  transformations on $A$ and $B$ which  are  compatible 
with  the  superpositions \eqref{E39}  and \eqref{E42}.    In fact,   these 
transformations  form a group that was discovered in  another 
context  by  Nishijima \cite{R33} and which has the following  explicit 
representation 
\begin{equation}\label{E43}
     G_N  = \sigma \exp(i\alpha) + \CON{\sigma} \exp(i\beta).   
\end{equation}
By  direct  calculation,  one finds that while $A$ and $B$ transform under 
$G_N$, $D$ transforms as $\exp(-\vec{\nu} \alpha)$
   and $N$ as $\exp(-\vec{\nu} \beta)$, 
respectively,  so that the system \eqref{E38} describes two particles of 
equal  mass but different  electric charges,  such as the  proton 
and the neutron.  Hence, by just trying to write Dirac's equation 
in  quaternions,   one  is  automatically  led  to  discover  the 
existence of \emph{isospin},  a fundamental feature that indeed is found 
in nature.\footnote{For more details, see Sec.~7 of Ref.~\cite{R35}.}

\section{Proca's equations and the absence of\\ magnetic monopoles} 
\label{pro:0}

     When  we wrote Maxwell's equation \eqref{E31} we made the  implicit 
assumptions $A=A^\REV$, i.e.,  that  $A$  was  a ordinal invariant 
fundamental  four-vector.   If we try now to put a mass term on  the 
right  of the second Maxwell equation,  and therefore introduce a 
``feedback''  to get the wave equation for a massive spin 1  field, 
we find  that  Maxwell's equations have necessarily to be  generalized  to  the 
following form
\begin{equation}\label{E44}
\left.
\begin{array}{c}
\CON{\nabla} \vect A  =  \vec{F}, \\
\tfrac{1}{2} (\nabla \vec{F}  + \vec{F}^\REV \nabla) = m^2  A.
\end{array}
\quad \right\}
\end{equation}
This  is,  written  in  biquaternions,  the correct spin  1  wave 
equation  discovered in  1936  by Proca \cite{R34}.   As  with  Dirac's 
equation it is easy to write in quaternions the conserved current  
and  the energy-momentum tensor
\begin{equation}\label{E45}
              2  J  =  A^+ \vec{F}  +  \vec{F}^+ A  +  (...)^\REV,  
\end{equation}
\begin{equation}\label{E46}
   8 \pi T\A = \vec{F}^+\Q \vec{F} + m^2 A^+\CON{\Q} A + (...)^\REV, 
\end{equation}
where $(...)^\REV$ means that  the expression  has to be completed  by 
adding the reverse of the part on the left.  Hence, the 
current  and the energy-momentum are bireal and ordinal invariant 
four-vectors, as it should be.

       Now,  just  as we derived Dirac's equation  from  Lanczos's 
equation \eqref{E38} by making the superposition \eqref{E39},  Proca's equation 
\eqref{E44}  can also be derived from \eqref{E38} by adding the second Lanczos 
equation to its reverse equation.  If this is so,  what then  is 
the  meaning  of the equation obtained by subtracting  Lanczos's 
second  equation from its reverse,  assuming  that  the 
potential $A$ is ordinal invariant
\begin{equation}\label{E47}
             \nabla \vec{F} -  \vec{F}^\REV \nabla  = 0    ~~~?    
\end{equation}
Obviously,  this  is just the part of Maxwell's  equation  which 
specifies  that  there  are \emph{no} magnetic  monopoles!   Hence,  if 
{Lanczos's} system \eqref{E38} is taken as the \emph{fundamental equation} from 
which  Dirac's  and  Proca's  equations  are  derived,  Maxwell's 
equation  is obtained by taking the $m^2 =0$ limit in \eqref{E44},  and 
\eqref{E47} insures the absence of magnetic monopoles.

However, if we would have assumed that  $A=-A^\REV$ instead of  $A=+A^\REV$, we would have found another fully covariant field equation, only differing from Proca's by the fact that a minus-sign would replace the plus-sign in \eqref{E44}: In fact, the correct equation for a massive pseudo-vector particle.  Therefore, as shown by G\"ursey in his PhD thesis \cite{R29}, the  wave-equations  of all scalar  and  vector particles, and of all pseu\-do-scalar  and  pseudo-vector particles, are just degenerated cases of Lanczos's fundamental equation \eqref{E38}.

\section{Einstein-Mayer: Electron-neutrino doublets\\ in 1933!}
\label{ein:0}

     When he wrote his 1929 papers on Dirac's  equation,  Lanczos 
was with Einstein in Berlin.  In 1933,  Einstein and Mayer (using 
semi-vectors,  a  formalism allied to quaternions) derived a spin 
$\tfrac{1}{2}$  field  equation (in fact, a generalized form of Lanczos's 
equation)  predicting  that particles would come in  doublets  of 
different  masses  \cite{R35,R36}.   The  idea was that  the  most  general 
Lagrangian for quaternionic fields, to be called the Einstein-Mayer-Lanczos (EML) Lagrangian, should have the form
\begin{equation}\label{E48}
 L = \Scal \Bigl[ A^+\CON{\nabla}A + B^+{\nabla}B  - (A^+ BE^+ + B^+ AE ) + (...)^+  \Bigr].  
\end{equation}
where $E \in \bbB$.  The field equations are then
\begin{equation}\label{E49}
\left.
\begin{array}{c}
\CON{\nabla} A = m B E^+, \\    
\nabla B = m A E.~~
\end{array}
\quad \right\}
\end{equation}
which reduce to \eqref{E38} when $E=m$.  In the general case,  the  second 
order  equations  for $A$ or $B$ become eigenvalue equations for  the 
\emph{mass} (the $m$ factor appearing  in  the argument of  $\exp \bigl(im(Et-\vec{p}\cdot\vec{x})\bigr)$ of plane wave solutions. 
This generalization is obtained by the substitutions 
$Am \rightarrow AE$ and $Bm \rightarrow Bm$ in the Lagrangian leading to \eqref{E38}.  Therefore, mass-generation is linked to a maximally parity violating field.

     There  are  two basic conserved  currents:  The  probability 
current $J$, and the barycharge current $K$
\begin{equation}\label{E50}
  J = AA^+   + \CON{BB^+} ~~~, ~~~~ K = AEA^+  + \CON{BEB^+}. 
\end{equation}
Keeping $E$ constant, $J$ is invariant in any non-Abelian unitary 
$SU(2) \otimes U(1)$
gauge transformation of $A$ or $B$.\footnote{In biquaternions, these transformations have the form $G\A = \Q e^{i\phi}\exp(\tfrac{1}{2}\theta\vec{a})$~.} On the other 
hand, $K$ is only invariant for Abelian gauge transformations  which 
also commute with $E$, i.e., elements of the general Nishijima group \eqref{E43} such that $E$ and $\sigma$ commute. 

     Of  special  interest  are the cases in which $E$ is  also  a 
global gauge field. The first such gauge is when $E$ is idempotent. 
One solution of \eqref{E49} is then massive and the other one  massless: 
An  electron-neutrino  doublet!  The  most  general  local  gauge 
transformations  compatible  with \eqref{E50} are then elements of the  unitary  Nishijima 
group  $ U_N(1,\bbC)$ combined with \emph{one} non-Abelian gauge transformation  which 
operates on either $E$ and $A$, or $E$ and $B$, exclusively.  This  leads 
directly to the Standard model of electro-weak interactions \cite{R37}.

     The second fundamental case is when $E$ is real: $E = E^*$. The eigenvalue equation is degenerate and the two masses are equal.  As shown by G\"ursey \cite{R32}, 
equation \eqref{E49} describes a nucleonic field,  non-locally coupled to 
a pseudoscalar sigma-pion field.

     Again, as in the examples shown in the previous sections, Hamilton's  
conjecture seems to be realized.  Einstein-Mayer's generalization \eqref{E49} lifts 
the  mass degeneracy of Lanczos's original equation \eqref{E38} and leads 
to  electron-neutrino doublets and weak interaction on one  hand, 
and to proton-neutron doublets and strong interaction in the form 
of the well known charge-independent pion-nucleon theory,  on the 
other hand \cite{R27}.\footnote{(Note added in 2006.) Soon after the publication of Einstein and Mayer's papers, Pauli asked his assistant Valentine Bargmann to study them.  Bargmann found that the solutions of Eq.~\eqref{E49} consist of doublets of particles with different electric charges and masses, including particles of zero mass \cite{BARGM1934-}.  It is unfortunate that Pauli, who had postulated the existence of the neutrino in 1927, did not understand the importance of Einstein and Mayer's papers, which contain all the key ingredients of the contemporary Standard Model if $E$ is interpreted as the Higgs field \cite{R60}.}

\section{Petiau waves and the mass spectrum of\\ elementary particles}
\label{pet:0}

     One  of the central problems of contemporary physics is  the 
question  of the origin of the mass of the elementary  particles.  
As we have seen,  by replacing the mass term in Lanczos's equation 
by some biquaternionic parameter,  Einstein has been able to show 
that  elementary particles come in doublets of different  masses.  
In  fact,  in 1930 already,  Lanczos wondered whether a theory in 
which the mass term is replaced by a variable would not simultaneously solve  
the problem of mass quantization  and that of infinities in field theory
\cite{R38}.

     Well, as nobody after Einstein and Mayer seemed to have taken 
Lanczos's suggestion seriously, 
one had to wait until 1965 (about the time when Gell-Mann  
and Zweig proposed the idea of quarks)  for somebody  to 
reinvent the concept.   This year,  in complete independence from 
mainstream research,  the French physicist Gerard Petiau  wrote a 
system  of  equations which may precisely give a solution to  the 
problem of the mass of the electrons and quarks \cite{R39}.

     Although   Petiau  was  thinking  in  very  general   terms, 
considering  complicated couplings between particles  of  various 
intrinsic  spin,  it  is  easy  with  quaternions  to  write  his 
fundamental equation in the case of spin $\tfrac{1}{2}$ particles \cite{R35}.   It 
just amounts,  in the spirit of Lanczos's feedback idea, to adding 
a third equation to Einstein's system \eqref{E49} in order to close it:
\begin{equation}\label{E51}
\left.
\begin{array}{c}
\CON{\nabla} A = B C, \\
\nabla  B = A C, \\
\nabla  C = A \CON{B}.
\end{array}
\quad \right\}
\end{equation}
Here $A$ and $B$ are the usual Lanczos spin $\tfrac{1}{2}$  fields,  and  the 
scalar $C$ an additional Einstein-Mayer field of spin  0.   Because 
the system is now closed, it becomes non-linear, and the 
solutions  are  much more constrained than with any usual  linear 
type of wave equations.

     For  instance,  the  single-periodic \emph{de Broglie  waves} that 
quantum  mechanics associates with a particle become double-periodic
\emph{Petiau waves} \cite{R40}.  Instead of being linear 
combinations 
of $\sin(z)$ and $\cos(z)$ functions,  these  waves are  superpositions 
of  elliptic functions ~$\operatorname{sn(z,k)}$,  ~$\operatorname{cn(z,k)}$,  etc.  A very appealing 
feature  of Petiau waves is that their dependence on the  modulus 
interpolates  between  pure de Broglie waves (for $k=0$)  and  pure 
solitonic  waves  (for $k=1$):  A  beautiful  realization  of  the 
wave/particle duality of quantum mechanics.  Moreover,  both  the 
amplitudes  and  the \emph{proper mass} (the $\mu$ factor appearing  in  the 
argument of  ~$\operatorname{sn} \bigl(\mu(Et-\vec{p}\cdot\vec{x}),k\bigr)$, for example) will be quantized.

     The most interesting thing,  however,  is happening when, in 
order  to  quantize  the  system,  the  Hamiltonian  function  is 
constructed.   Taking,  for  example, $A$ as the fundamental field 
Petiau  showed that,  in terms of the first  integrals,  the 
Hamiltonian has the very simple form \cite{R39}
\begin{equation}\label{E52}
                           H = C_0 \mu^4 k^2,              
\end{equation}
where $k$ is the modulus of the elliptic function, $\mu$ the  proper 
mass,  and $C_0$ some constant.   The exciting thing  is  that  the 
Hamiltonian,  and thus the total energy in the field (i.e., for a 
single particle, the \emph{effective mass}) scales with the fourth power 
of $\mu$.

     In  effect,  in  1979,  Barut  discovered a  very  good 
empirical formula for the mass of the leptons \cite{R41}.  Assuming that a 
quantized self-energy of magnitude  $\frac{3}{2} \alpha^{-1} M_e c^2 N^4$, where 
$N = 0, 1, 2, ...,$  is  a new quantum number,  be added to the rest-mass  of a 
electron  to  get the next heavy lepton in  the  chain  $e$, $\mu$, 
$\tau$, ..., Barut got the following expression (where $\alpha = 1/137$)
\begin{equation}\label{E53}
 M(N)= M_e (1+\tfrac{3}{2}\alpha^{-1} \sum_{n=0}^{n=N} n^4). 
\end{equation}
The  agreement   with the data of this rather simple  formula  is 
surprisingly good, the discrepancy being of order $10^{-3}$ for $\mu$ and 
$10^{-3}$  for $\tau$,  respectively \cite{R42}.   In order to get the masses of 
the  quarks  \cite{R43},  it  is enough to take for  the  mass  of  the 
lightest quark $M_u = M_e/7.47$~. Again, as can be seen in {\bf Table 1} the
agreement between the theoretical quark masses and the ``observed'' 
masses is quite good, especially for the three heavy quarks.

     Since  we  have just seen \eqref{E52} that the energy of  a  Petiau 
field  is  scaling with the fourth power,  we are inclined to  think 
that  there might indeed be a fundamental link between such  non-linear fields and the theory of the mass of quark and  electrons.  
If this is so, what about the factor 7.47 ?

     There  are  two non-trivial exceptional cases  for  elliptic 
functions:  The harmonic  case, $k = \sin(\frac{\pi}{4})$, and the 
equianharmonic case, $k = \sin(\frac{\pi}{12})$.  It is very plausible  to 
associate  the former with leptons,  and the latter with  quarks.  
Indeed,  in  either case,  the corresponding  elliptic  functions 
exhibit  several  unique symmetry and scaling  properties,  which 
come  from the fact that in the complex plane their poles form  a 
modular aggregate with  $\frac{\pi}{2}$ or $\frac{\pi}{3}$ symmetries.  Since 
according to 
(52) the masse is proportional to $k^2$,  the electron to quark mass 
ratio is then equal to $[ \sin(\frac{\pi}{4}) / \sin(\frac{\pi}{12}) ]^2 \approx 7.47$ .

     But this is now very close to pure speculation,  and in  any 
case    on  the  frontier  of  contemporary   research   \cite{R43}.  
Nevertheless,  it is interesting to see how far, just following 
Hamilton's  conjecture,  one  can  go in the  direction of a  
unified picture of fundamental physics.

\begin{center}
                           {\bf \Large Table 1 }
\end{center}

\begin{tabular}{|c|c|c|c|c|c|c|}
	\hline
	N & \multicolumn{3} {c|} {electron masses} 
	  & \multicolumn{3} {c|} {quark masses}  \\
	\hline
	  &   & Barut's  &          &   & Barut's &          \\
	
	  &   & formula  & Data     &   & formula & Data     \\
	\hline
	0 & e & \hskip1.4cm   0.511 & \hskip 1.0 cm 0.511 & u & \hskip1.3cm 0.068 &   0 -- 8  \\
	
	1 &$\mu$& \hskip0.8cm 105.55 & \hskip 0.4 cm 105.66 & d & \hskip0.7cm14.1 &   5 -- 15 \\
	
	2 &$\tau$& \hskip0.4cm 1786.1 &   1784.1 & s & \hskip0.4cm 239  & 100 -- 300 \\
	
	3 &   &   10294  &     ?    & c & \hskip0.2cm 1378   & 1300 --1500 \\
	
	4 &   &   37184  &     ?    & b & \hskip0.2cm 4978   & 4700 -- 5300 \\
	
	5 &   &          &          & t &  13766  &      ?    \\
	
	6 &   &          &          &   &  31989  &      ?    \\
	\hline

\end{tabular}

\noindent \emph{Comparison of electron and quark masses in MeV/c$^2$ calculated 
with Barut's formula \eqref{E53} to measured masses from Ref.~\cite{R42}. (Note added in 1996:  The observation of a sixth quark of mass in the range of 160'000 to 190'000 MeV/c$^2$ has been reported at the beginning of 1995.)}

\section{Quaternions and quantum mechanics}
\label{qua:0}

     Some  of  those who have been following us on  this  upwards 
trail,  starting  from  Hamilton's  principle and ending  with  a 
possible  solution  to  the problem of  quark  masses,  might  be 
surprised  that  just  a short section is  dedicated  to quaternions and
quantum mechanics.   To these we say --- in the  spirit  of Hamilton's 
particle-wave duality --- that everything we have done  can easily 
be recast in the jargon of ``wave mechanics,'' so that, in this perspective, 
we  have been  doing quantum mechanics all along.

     In effect,  the quantum predicate is no so much in the field 
equations  we have been discussing in this paper,  than in 
the   interpretation  of  the  field's  amplitudes   (the   ``wave 
functions:''   complex   number  in  the  non-relativistic   case, 
biquaternions  in  the  relativistic one).   As  Feynman  clearly 
stated  in a review of the principles of quantum  mechanics:  ``It 
has  been  found  that  all processes  so  far  observed  can  be 
understood  in  terms  of the following  prescription:  To  every 
\emph{process} there corresponds an \emph{amplitude}; with proper 
normalization the probability of the process is equal to the absolute 
square of this amplitude'' Ref.~\cite{R44}, page 1.

     Take,   for  example,   the  Dirac-Lanczos  equation \eqref{E40}, 
rewritten   here  in  the  case  where  there  is   an   external 
electromagnetic field $A_{\text{em}}$
\begin{equation}\label{E54}  
        \CON{\nabla} D
               = (m D^* + e \CON{A}_{\text{em}} D) i \vec{\nu}.   
\end{equation}
Comparing with (1),  we see that the Hamiltonian is the following 
operator
\begin{equation}\label{E55}
      H\A = -\vec{\nabla}\Q  - m\Q ^*i \vec{\nu}
            - e \CON{A}_{\text{em}}\Q i \vec{\nu}. 
\end{equation}
Then, following Feynman's prescription, we have to normalize the 
amplitude $D$.  Since the probability 
current is the conserved four-vector $DD^+$, a suitable norm is
\begin{equation}\label{E56}
 \BRA D |\mathbf{1}| D \KET ~ = ~ \iiint d^3V ~ \SCA D^+[\mathbf{1}]D \LAR ~ = 1, 
\end{equation}
where the dummy operator ``$\mathbf{1}$'' may be replaced by the
operator corresponding to the physical quantity whose  expectation value is 
to be calculated. Hence once the above prescription has been  accepted 
as  a postulate there is not much mystery left in quantum theory, 
and  it is straightforward,  at least in principle,  to give  the 
quantum  interpretation of the field equations presented  in  the 
previous sections.\footnote{It is remarkable that it is the use of complex conjugation in expressions such as the Hermitian product \eqref{E56} that distinguishes quantum theory from classical physics. }  

     At  this  point,  it is worth mentioning that in  the  past 
decade  a  kind  of a silent revolution has  been  occurring  with 
respect to quantum theory.   Increasingly,  quantum and classical 
theories  are seen as part of the same theory:  Barut shows  that 
wave mechanics can be formulated without $\hbar$ \cite{R45}, Lamb suggests 
that  Newton could have invented wave mechanics \cite{R46}, and  several 
major investigations show that the whole apparatus of the so-called  
``second  quantization''  of fields is  redundant  \cite{R47}.  
Indeed,  as is trivially shown in the way Lanczos's equation \eqref{E38} 
generalizes Maxwell's equation \cite{R31},  Maxwell's theory can be interpreted 
as a quantum theory without $\hbar$.  For instance, calculating  by 
means of Silberstein's energy-momentum tensor \eqref{E32} the electromagnetic 
field's energy-momentum density, or the Lorentz's force density \eqref{E33}, 
is the same as applying the quantum rule \eqref{E56}.  Moreover, 
the Hamiltonian is simply the operator 
$H\A  = \tfrac{1}{2}(\vec{\nabla}\Q  - \Q \vec{\nabla})$ just like in Good's 
quantum interpretation of Maxwell's theory \cite{R48}.

     What,  is then the main contribution of biquaternions to 
quantum theory?   Possibly,  the clear disentanglement of ``$i$'' and 
``$\hbar$,''  the  two elements which have been traditionally  associated 
with  quantum  mechanics.   Indeed,  looking at  the  Schr\"odinger 
equation \eqref{E1},  these two elements appear together as a  combined 
factor. In Lanczos equation \cite{R38}, however, ``$i$'' appears in the 
scalar part of the four-gradient $\nabla$, and $\hbar$ in combination 
with the mass on  the right-hand side had we not
used the convention $\hbar=1$.  Thus, if Hamilton's conjecture is true, 
``$i$'' is definitely associated with ``time'' (i.e., Hamilton's intuitive 
conception of imaginary numbers) and ``$\hbar$'' is associated with ``mass'' 
or, more precisely, with the particle aspect of waves, i.e., lumps 
of energy localized in space \cite{R49,R43}.\footnote{(Note added in 2006.) The origin of ``$i$'' in quantum theory is explained by Lanczos as follows: ``For reasons connected with the imaginary value of the fourth Minkowskian coordinate $ict$, the wave mechanical functions assume \emph{complex values}'' \cite[p. 268]{LANCZ1970-}.}

     It  remains,  in  conclusion,  to stress that the  power  of 
Hamilton's conjecture seems not to have shown its limits yet.  By 
this  we  allude to the numerous investigations in  ``quaternionic 
quantum  mechanics''  which have occurred since the birth  of  wave 
mechanics.

     Indeed,   not   to  mention  the  work  of  Lanczos \cite{R26,R38}, 
quaternionic  and other more general algebraic generalization  of 
quantum mechanics have been actively studied since 1928 already \cite{R50}. The 
best  known  sequel  to this work is possibly  what  Jordan  \cite{R51} 
started  in 1932 and which culminated in the famous article  \cite{R52} 
in  which  the  first ``Jordan algebra''  was  described.  To  give 
another example of the breadth of research in these early days,  we 
mention that the theory of operators in quaternionic Hilbert spaces 
was the subject of a PhD thesis in 1935, and that the name ``Wachs 
space'' was proposed for such spaces \cite{R53}.

     A new impetus was given in the 1960's, mainly after the work  
of the group around Finkelstein and Jauch \cite{R54} at CERN,  followed 
by others \cite{R55}, up to the synthesis soon to be published by Adler 
\cite{R56}. All these developments contemplate the possibility that the 
complex  numbers  of contemporary quantum theory may have  to  be 
replaced by quaternions or biquaternions in some more fundamental 
theory.  However, it  may well be that Nature is satisfied with complex 
number as the fundamental scalar field,  and that in this respect 
a single and commutative ``$i$'' is enough and playing some essential role
that the current experimental situation seems to favor \cite{R57,R58}.

\section{Conclusion}
\label{con:0}

The physics is in the mathematical \emph{structure}, not in the \emph{formalism}: What are then the advantages of using a formalism such as Hamilton's biquaternions?
\begin{itemize}

\item Biquaternions are {\it whole symbols}, i.e., they compound between one and eight real numbers which belong to a single (or a few related) tensor quantity(ies) so that many formulas written in biquaternions are simpler than their standard matrix, tensor, or higher rank Clifford numbers counterparts. In general, they enable to dispense of at least one level of tensor indices, and quite often to reduce a few indices tensor to a single entity.
  
\item Biquaternions are {\it expressive}, i.e., being elements of the simplest non-trivial Clifford algebra they provide neat and  explicit 
formulas in final form,  which are therefore directly amenable to symbolic or numerical calculation, with pencil and paper, or with a computer.

\item Biquaternion formulas are {\it suggestive}, i.e., they often indicate the correct way of generalizing a result, or how to relate seemingly independent results.


\item Biquaternions provide a {\it unifying formalism,} e.g., they enable a fully consistent use of complex numbers in both classical and quantum physics; they lead to expressions that are very similar in both Galilean and Lorentzian relativity; they are very effective in formulating the physics of the current ``Standard model'' of fundamental interactions \cite{R27,R35,R60}; etc.


\end{itemize}
\noindent However,  some words of caution are in order: Since biquaternions are whole symbols it is important to take care of the problems specifically associated with such symbols. For instance, in order that biquaternion expression have a well defined tensor character they have to be constructed from elementary biquaternions that have such a character.  Moreover, special care is required because of noncommutativity and of the need for biquaternion expression to be {\it ordinal covariant.}  For example, the truly correct form of Maxwell's equations is not the Conway-Silberstein expression \eqref{E31}, but the gauge and ordinal invariant system
\begin{equation}\label{E57}
\left.
\begin{array}{c}
\CON{\nabla} \vect A  =  \vec{F}, \\
\tfrac{1}{2} (\nabla \vec{F}  + \vec{F}^\REV \nabla) = -4 \pi  J,
\end{array}
\quad \right\}
\end{equation}
which like Proca's equation \eqref{E44} does not require the supplementary conditions $\SCA\CON{\nabla}A\LAR=0 , A=A^\REV , J=J^\REV$ , i.e., the Lorenz gauge and the requirements that $A$ and $J$ are ordinal invariant biquaternions.  It is is precisely because such problems were not properly understood and cured that biquaternions failed, at the beginning of the twentieth century, to become a widespread language for physics.

\section{Acknowledgments}
\label{ack:0}

     We  wish  to thank Professor  Brendan  Goldsmith,  Principal 
Organizer of the Conference commemorating the sesquicentennial of 
the  invention of quaternions,  for inviting us to give a lecture 
on  the  Physical heritage of Sir  W.R.  Hamilton,  although  the 
audience was primarily composed of mathematicians.  We also wish 
to thank Professor James R.  McConnell for his encouragement  and 
suggestion  to  present  a  more  technical  account  of  our 
findings   at  the  Cornelius  Lanczos  International   Centenary 
Conference \cite{R27}.  And, finally, we are  indebted to Mrs Ann Goldsmith, 
Librarian  at the Dublin Institute of Advanced Studies,  for  her 
kind  assistance  in helping us retrieving some  rather  old  but 
nevertheless important quaternion references \cite{R59}.


\section{References}
\label{bib:0}


\begin{enumerate}

\bibitem{R1}  Most   biographies  of  Hamilton give some  details  on his 
conception  of algebra as the science of pure  time.  See   e.g., 
O'DONNELL  S.  (1983).  William  Rowan  Hamilton  --- Portrait  of  a 
Prodigy.  Bool Press. Dublin. See also \cite{R2}.

\bibitem{R2}  LANCZOS  C.  (1967). \emph{William Rowan Hamilton --- An  appreciation}, Amer. Sci.,  {\bf 55}, 129--143. Reprinted  {\bf in} DAVIS W.R. \emph{et al.}, eds., Cornelius Lanczos Collected Published Papers With Commentaries (North Ca\-rolina State University, Raleigh, 1998) Vol. IV pages 2-1859 to 2-1873.
  
\bibitem{R3}  GOUGH W. (1990). \emph{Mixing scalars and vectors --- an elegant view of physics},  Eur. J. Phys.  {\bf 11}, 326--333.  

\bibitem{R4}   WIGNER  E.P.   (1960). \emph{The  unreasonable   effectiveness   of mathematics in the natural sciences},  Comm.  Pure and Appl. Math.  
{\bf 13}, 1--14.

\bibitem{R5} GURSEY F. (1956), \emph{Contribution to the quaternion formalism in special relativity}, Rev. Fac. Sci. Istanbul {\bf A 20} 149--171.  G\"ursey uses both ``quaternion units'' $\{1, e_1, e_2, e_3\}$ and ``Hermitian units'' $\{1, I_1, I_2, I_3\}$ with $I_n=ie_n$.  See also \cite{R24}.

\bibitem{R6}  E.C.G. St\"uckelberg et al., Helv. Phys. Acta. {\bf 33}, 727--752 (1960); {\bf 34}, 621--628 (1961).  See also section 2.6 of Ref.~\cite{R56}, pages 47--49.

\bibitem{R7}  HAMILTON W.R.  (1866). Elements of Quaternions.  Second 
edition 1899-1901  enlarged by C.J.  Joly.  Reprinted in 1969 by Chelsea 
Publishing, New York.

\bibitem{R8} SYNGE J.L. (1972). Quaternions,   Lorentz  transformations  and  the   Conway-Dirac-Eddington matrices.   Comm.  Dublin Inst. for Adv. Studies.  Series A. No. 21.

\bibitem{R9} BLATON J.  (1935). \emph{Quaternionen, semivektoren and spinoren},  Z. f. Phys.  {\bf 95}, 337--354.

\bibitem{R10}  See,  for  example,  chapter  IX of  LANCZOS  C.  (1986). The 
Variational Principles of Mechanics.  Dover Publ. New York.

For two other good introductions to the use of quaternions in special 
relativity, see \cite{R5,R8}.

\bibitem{R11} LANCZOS C. (1958) \emph{Tensor calculus}, {\bf in}  CONDON E.U. and ODISHAW H., eds., Handbook of Physics. (McGraw-Hill, New York) Part I., Ch. 10, I-140 to 1-151.  Reprinted  {\bf in} DAVIS W.R. \emph{et al.}, eds., Cornelius Lanczos Collected Published Papers With Commentaries (North Ca\-rolina State University, Raleigh, 1998) Vol. IV pages 2-1752 to 2-1763.

\bibitem{R12} RASTALL P. (1960). \emph{Quaternions in relativity},  Rev. Mod. Phys. {\bf 36}, 820-832.

See also \cite{R29} in which G\"ursey generalizes Lanczos's fundamental equation to curved space-time.

\bibitem{R13}  CONWAY A.W. (1911). \emph{On the applications of quaternions to some recent developments of electrical theory},  Proc. Roy. Irish Acad.  {\bf 29}, 1--9.

\bibitem{R14} SILBERSTEIN L. (1912). \emph{Quaternionic form of relativity}, Phil. Mag. {\bf 23}, 790--809.
 
CONWAY A.W.  (1912). \emph{The quaternion form of relativity}, Phil.  Mag.  
 {\bf 24}, 208.

\bibitem{R15}   SILBERSTEIN   L.   (1913). \emph{Second  memoir  on   quaternionic relativity}, Phil. Mag.   {\bf 25}, 135--144.

\bibitem{R16}  JACKSON J.D.  (1975). Classical Electrodynamics.    Wiley.  New 
York.

\bibitem{R17}  KILMISTER  C.W.  (1955). \emph{The application  of  certain  linear quaternion functions to tensor analysis}, Proc.  Roy. Irish Acad.   {\bf 57}, 
37-99.

\bibitem{R18} LANCZOS C.  (1919). Die Funktionentheoretischen Beziehungen 
der Max\-well\-schen  Aethergleichungen --- Ein Beitrag  zur  Relativit\"ats- und  Elektronentheorie. Verlagsbuchhandlung Josef N\'emeth. Budapest. Reprinted  {\bf in} DAVIS W.R. \emph{et al.}, eds., Cornelius Lanczos Collected Published Papers With Commentaries (North Ca\-rolina State University, Raleigh, 1998) Vol. VI pages A-1 to A-82.

Lanczos's dissertation is available in typeseted form as: 

LANCZOS C.  (1919) \emph{The relations of the homogeneous Maxwell's equations to the theory of functions --- A contribution to the theory of relativity and electrons}. Typeseted by Jean-Pierre Hurni with a preface by Andre Gsponer, 2004, 58~pp.  e-print \underline{ arXiv:physics/0408079 }. 

\bibitem{R19} GSPONER A.  and HURNI J.P. (1994). \emph{Lanczos's functional theory of electrodynamics},  {\bf in} DAVIS W.R. \emph{et al.}, eds., Cornelius Lanczos Collected Published Papers With Commentaries (North Ca\-rolina State University, Raleigh, 1998) Vol. I pages 2-15 to 2-23. e-print \underline{ arXiv:math-ph/0402012 }.

See also:

GSPONER A.  and HURNI J.P. (2005). \emph{Cornelius Lanczos's derivation of the usual action integral of classical electrodynamics}, Foundations of Physics {\bf  35}, 865--880. e-print \underline{ arXiv:math-ph/0408100 }.
 
\bibitem{R20}  FUETER R.  (1932). \emph{Analytische Funktionen einer  Quaternionen-variablen}, Comm. Math. Helv.   {\bf 4}, 9--20.

MOISIL C.R. (1931). \emph{Sur les quaternions monog\`enes},  Bull. Sci. 
Math.  {\bf  55}, 168--174.

For a bibliography of Fueter's work, see:
HAEFELI H.G.  (1947). \emph{Hypercomplex Differentiale}, Comm. Math. 
Helv.  {\bf  20}, 382--420.

For an elementary introduction to quaternion calculus, see:
DEAVOURS C.A.  (1973). \emph{The quaternion calculus},  Am.  Math. 
Monthly  {\bf  80}, 995--1008.

For a more professional introduction, see: SUDBERRY A.  (1979). \emph{Quaternionic analysis}, Math. Proc. Camb. Phil. 
Soc.  {\bf  85}, (199--225.

\bibitem{R21}  RYAN  J.   (1982). \emph{Complexified  Clifford  analysis}, Complex Variables   {\bf 1}, 119--149.

BRACKX  F.,  DELANGHE R.  and SOMMEN F.  (1982). Clifford  Analysis.  
Pitman Books. London. 307 pp.

SOUCEK V.  (1983). \emph{Complex-quaternionic analysis applied to spin-$\tfrac{1}{2}$ massless fields},  Complex Variables  {\bf 1}, 327--346.

\bibitem{R22}   IMAEDA   K.   (1976).  \emph{A  new   formulation   of   classical electrodynamics}, Nuovo Cim.  {\bf 32B}, 138--162.


\bibitem{R23} PROCA  A. (1955). \emph{Particules de tr\`es grande vitesse  en  
m\'ecanique spinorielle}, Nuovo Cim.  {\bf 2}, 962--971.

GURSEY  F.  (1957). \emph{Relativistic Kinematics of  a  Classical  Point 
Particle in Spinor Form}, Nuovo Cim.   {\bf 5}, 784--809.

PENROSE R. (1963). \emph{Null hypersurface initial data for classical fields  of arbitrary spin and for general  relativity}, Reprinted  in
1980  Gen. Rel. and Grav.  {\bf 12}, 225--264.


\bibitem{R24}  WEISS  P.  (1941).  \emph{On  some applications  of  quaternions  to restricted relativity and classical radiation  theory},  Proc.  R. 
Ir. Acad.   {\bf 46}, 129--168.

Despite  the internal consistency of this article,  we  recommend 
the  readers not to use Weiss's ``Hermitian'' units,  but Hamilton's 
real quaternion units.

\bibitem{R25} DIRAC P.A.M.  (1938). \emph{Classical theory of radiating electrons}, Proc. Roy. Soc.  {\bf A167}, 148--169.

(Note added in 2006) Before Dirac this equation was derived by Myron Mathisson using a more satisfactory method than Dirac's.  See:

MATHISSON M. (1931). \emph{Die Mechanik des Materieteilchens in der allgemeine Relativit\"atstheorie}, Z.  f. Phys. {\bf 67} 826--844.



\bibitem{R26}  LANCZOS  C.  (1929). \emph{Die  tensoranalytischen  Beziehungen  der Diracschen Gleichung},  Z.  f. Phys.   {\bf 57}, 447--473, 474--483, 
484--493. Reprinted and translated {\bf in} DAVIS W.R. \emph{et al.}, eds., Cornelius Lanczos Collected Published Papers With Commentaries (North Ca\-rolina State University, Raleigh, 1998) Vol. III pages 2-1133 to 2-1225. e-prints \underline{ arXiv:physics/0508002 }, \underline{ arXiv:physics/0508012 }, and \underline{ arXiv:physics/0508013 }.

\bibitem{R27} GSPONER A. and HURNI J.P. (1994). \emph{Lanczos's equation to replace Dirac's equation?},  {\bf in} BROWN J.D. \emph{et al.}, eds., Proc. of the Cornelius Lanczos Int. Centenary Conf. December 12--17, 1993, Raleigh, NC, USA. (SIAM, Philadelphia, 1994) 509--512.  There are a number of typographical errors in this paper. Please refer to the e-print \underline{ arXiv:hep-ph/0112317 }.

\bibitem{R28}  EDMONDS  J.D.   Jr.   (1976).  \emph{A  relativistic  ``higher  spin'' quaternion  wave  equation  giving  a  variation  on  the   Pauli 
equation},   Found.  Phys.   {\bf 6}, 33--36; (1977  ibid  {\bf 7}, 835--859; 
1978  ibid  {\bf 8}, 439--444.

GOUGH W.  (1989). \emph{A quaternion expression for the quantum mechanical 
probability and current densities} Eur. J. Phys.   {\bf 10}, 188--193.

See also reference \cite{R60}.

\bibitem{R29}  GURSEY  F.  (1950). Applications  of  Quaternions  to  Field 
Equations.  PhD thesis. University of London. 204 pp.

In  his PhD thesis,  G\"ursey reviews the quaternionic theories  of 
LANCZOS, Ref.~\cite{R26}, and others, in particular:

CONWAY   A.W.  (1937). \emph{Quaternion Treatment of the relativistic Wave 
Equation}  Proc. Roy. Soc.   {\bf A162}, 145--154.

\bibitem{R30} HESTENES D.  (1967). \emph{Real spinor fields}, J. Math. Phys.  {\bf 8}, 798--808, 809--812. 

Following Hestenes,  there is  nowadays a renewed interest in the 
Clifford  formulation  of Dirac's theory,  a framework  that  was 
first independently developed by Sauter, Proca and Juvet:

SAUTER   F.   (1930). \emph{L\"osung  der Diracschen   Gleichungen   ohne 
Spezialisierung der Diracschen Operatoren},  Z. f. Phys.   {\bf 63}, 803--814;  ibid,  {\bf 64}, 295--303.

PROCA A.  (1930). \emph{Sur l'\'equation de Dirac},  J.  Phys. Radium   {\bf 1}, 235--248.

JUVET G.  (1930). \emph{Operateurs de Dirac et \'equations de Maxwell}, Comm. 
Math. Helv.  {\bf 2} 225--235.
 
\bibitem{R31}  GURSEY F.  (1958). \emph{Relation of charge independence and  baryon conservation to Pauli's transformation}, Nuovo Cim.   {\bf 7}, 411--415.

\bibitem{R32}  GURSEY  F.  (1960). \emph{On  the symmetries  of  strong  and  weak interactions},  Nuovo Cim.   {\bf 16}, 230--240.

GELL-MANN M.  and  LEVY  M. (1960). \emph{The axial vector current in beta 
decay},  Nuovo Cim.   {\bf 16}, 705--725.

\bibitem{R33} NISHIJIMA K.  (1957). \emph{On the theory of leptons},  Nuovo Cim. {\bf 5}, 1349--1354.

\bibitem{R34}  PROCA  A.  (1936). \emph{Sur la th\'eorie ondulatoire  des  \'electrons positifs et n\'egatifs},  J. Phys. Radium   {\bf 7}, 347--353.

CONWAY A.W. (1945). \emph{Quaternion and matrices},  Proc. Roy. Irish Acad.  
 {\bf A50}, 98--103.

\bibitem{R35}  GSPONER A. and HURNI J.P. (1994). \emph{Lanczos-Einstein-Petiau: From Dirac's equation to non-linear wave mechanics}, {\bf in} DAVIS W.R. \emph{et al.}, eds., Cornelius Lanczos Collected Published Papers With Commentaries (North Ca\-rolina State University, Raleigh, 1998) Vol. III pages 2-1248 to 2-1277.  e-print \underline{ arXiv:physics/0508036 }.

\bibitem{R36}  EINSTEIN  A.  and MAYER W.  (1933).  \emph{Die  Diracgleichung  f\"ur Semivektoren}, Proc. Roy. Acad. Amsterdam   {\bf 36}, 497--516, 615--619. 
For the idea of replacing the mass by a biquaternionic parameter, 
see also EDMONDS, Ref.~\cite{R28}.

\bibitem{R37}  BJORKEN  J.D.  (1979). \emph{Neutral-current results without  gauges theories}, Phys. Rev.   {\bf D19}, 335--346.

\bibitem{BARGM1934-}  BARGMANN V.  (1934), \emph{Uber den Zusammenhang zwischen Semivektoren und Spinoren und die Reduktion der Diracgleichungen f\"ur Semivektoren}, Helv. Phys. Acta {\bf 7} 57--82.

\bibitem{R38}  LANCZOS  C.  (1930). \emph{Dirac's  wellenmechanische  Theorie  des Elektrons  und  ihre  feldtheoretische   Ausgestaltung}, Physik. 
Zeitschr.   {\bf 31}, 120--130. Reprinted and translated {\bf in} DAVIS W.R. \emph{et al.}, eds., Cornelius Lanczos Collected Published Papers With Commentaries (North Ca\-rolina State University, Raleigh, 1998) Vol. III pages 2-1226 to 2-1247. e-print \underline{ arXiv:physics/0508009 }.

\bibitem{R39}  PETIAU  G.  (1965).  \emph{Sur les th\'eories  quantiques  des  champs associ\'es   \`a   des  mod\`eles  simples  d'\'equations   d'ondes   
non lin\'eaires}, Nuovo Cim.   {\bf 40}, 84--101. For a synthesis of Petiau's work on this subject,  see: PETIAU   G.   (1982). \emph{Sur  la  repr\'esentation  par  des   syst\`emes diff\'erentiels  du premier ordre de mod\'eles  corpusculaires},  Ann. Inst. Henri Poincar\'e   {\bf 36}, 89--125.

\bibitem{R40}  PETIAU  (1958). \emph{Sur une g\'en\'eralisation  non  lin\'eaire  de  la m\'ecanique ondulatoire et sur les propri\'et\'es des fonctions 
d'ondes correspondantes}, Nuovo Cim.   {\bf 9}, 542--568.

\bibitem{R41} BARUT A.O.  (1979). \emph{Lepton mass formula},  Phys.  Rev. Lett. {\bf 42}, 1251.

\bibitem{R42}  PARTICLE   DATA GROUP (1992). \emph{Review of  Particle  Properties}, Phys. Rev.   {\bf D45}, II-4. According to this reference,  page II-4, note [g], ``the estimates 
of $d$ and $u$ masses are not without controversy ...  the  $u$-quark 
could be essentially massless.''  See also GASSER J. and LEUTWYLER 
H. (1982). \emph{Quark Masses}, Phys. Rep.   {\bf 87}, 77--169.

\bibitem{R43} GSPONER A. and HURNI J.P. (1994). \emph{A non-linear field theory for the  mass of the electrons and quarks}, Hadronic Journal {\bf19} (1966) 367-373.  e-print \underline{arXiv:hep-ph/0201193  }. After this paper was published we learned from G. Rosen that  H. Terazawa also noted that the $n^4$ proportionality could be extended to quarks on page 1769 of TERAZAWA, H. (1980) Prog. Theor. Phys. {\bf 64} 1763--1771. 

\bibitem{R44}  FEYNMAN  R.P.  (1962). The Theory  of  Fundamental  
Processes.  Benjamin. Reading, MA, USA.

\bibitem{R45}  BARUT A.O.  (1992). \emph{Formulation of wave mechanics without  the Planck constant $\hbar$},  Phys. Lett.  {\bf  A171}, 1--2

\bibitem{R46}  LAMB  W.E.  Jr.  (1994).  \emph{Suppose  Newton  had  invented  wave mechanics}, Am. J. Phys.   {\bf 62}, 201--206.

\bibitem{R47} SCHWINGER J. (1969, 1973, 1988).  Particles, Sources and Fields.  
3 volumes. Addison-Wesley. Reading, MA.

BARUT    A.O.     (1990).  \emph{Foundations    of    self-field 
quantum electrodynamics},  {\bf in}   BARUT  A.O.  Ed.   New  Frontiers  
in Quantum electrodynamics  and  Quantum Optics.   Plenum  Press.  New 
York.

\bibitem{R48}   GOOD R.H. (1957). \emph{Particle aspect of the electromagnetic field equations}, Phys. Rev. {\bf 105}, 1914--1919.
 

\bibitem{R49} BARUT A.O. (1990). $E = \hbar \omega$.  Phys. Lett.   {\bf A143}, 349--352.

\bibitem{LANCZ1970-}  LANCZOS C. (1970) Space Through the Ages --- The Evolution of Geometrical Ideas from Pythagoras to Hilbert and Einstein. Academic Press. New York.

\bibitem{R50}  GURSEY  F.  (1983). \emph{Quaternionic and octonionic structures  in physics}, Proc. of the 1st. Int. Meeting on the Hist. of Scientific 
Ideas.  Barcelona.  29, 559--591.

\bibitem{R51}  JORDAN P.  (1933). \emph{\"Uber  Verallgemeinerungsm\"oglichkeiten  des Formalismus Quantenmechanik},  Nachr.  Ges.  Wiss.   G\"ottingen   {\bf 39}, 
209--217.

\bibitem{R52} JORDAN P., VON NEUMANN J. and WIGNER E. (1934). \emph{On an 
algebraic generalization of the quantum mechanical formalism}, Ann. Math.  
 {\bf 35}, 29--64.

\bibitem{R53}  TEICHMULLER O.  (1935). \emph{Operatoren in Wachschen Raum}, J.  
f\"ur Math.  {\bf 174}, 73--124

\bibitem{R54}  FINKELSTEIN D.,  JAUCH J.M.  and SPEISER D.  (1959). \emph{Notes  on quaternion quantum mechanics}, CERN report 59-7. Published {\bf in}  
HOOKER C.A.,  ed., Logico-Algebraic   Approach  to  Quantum  Mechanics.  (1979, Reidel, Dordrecht). Vol. II  367--421.

\bibitem{R55} In particular: 

RUELLE D. (1958). \emph{Repr\'esentation du spin isobarique des particules \`a 
interaction forte},  Nucl. Phys.   {\bf 7}, 443--450.

KANENO T. (1960). \emph{On a possible generalization of quantum mechanics}, 
 Progr. Theor. Phys.   {\bf 23}, 17--31.     

ADLER  S.L.   (1980).  \emph{Quaternion  chromodynamics  as  a  theory  of 
composite quarks and leptons},  Phys. Rev.   {\bf D21}, 2903--2915.

HORWITZ  L.P.   and  BIEDENHARN  L.C.   (1984). \emph{Quaternion  quantum 
mechanics:  Second quantization and gauge fields}, Ann. Phys.  
 {\bf 157}, 432--488.

NASH C.G.  and JOSHI G.C.  (1987). \emph{Composite systems in quaternionic 
quantum mechanics},  J. Math. Phys.   {\bf 28}, 2883--2885.

ROTELLI P.  (1989). \emph{The Dirac equation on the quaternion field},  Mod. 
Phys. Lett.   {\bf A4}, 933--40.

SHARMA  C.S.  and  ALMEDIA D.F.  (1990). \emph{Additives isometries  on  a 
quaternionic Hilbert space},  J. Math. Phys.   {\bf 31}, 1035--1041.

NASH C.G.  and JOSHI G.C.  (1992). \emph{Quaternionic quantum mechanics is 
consistent with complex quantum mechanics},  Int.  J. Theor. Phys.  
 {\bf 3}, 965--981

MARCHIAFAVA S.  and REMBIELINSKI J (1992). \emph{Quantum  quaternions}, 
 J. Math. Phys.   {\bf 33}, 171--173.

\bibitem{R56}  ADLER  S.L.  (1995). Quaternionic  Quantum  Mechanics.   
Oxford University Press. London.

\bibitem{R57}  PERES A.  (1979). \emph{Proposed test for complex versus quaternion quantum theory}, Phys. Rev. Lett.   {\bf 42}, 683--686.

Note, in particular, Ref.~\cite{R18} cited therein.

\bibitem{R58}  DAVIES  A.J.   and  McKELLAR  B.H.  (1992).  \emph{Observability  of quaternionic quantum mechanics}, Phys. Rev.   {\bf A46}, 3671--3675.

\bibitem{R59}  In  the  course  of  this  research  we  have  compiled  an 
essentially   complete   bibliography  on  the  applications   of 
quaternions in ``modern'' physics,  i.e.,  starting with relativity:

  
GSPONER A. and HURNI J.P. (1993). \emph{Quaternion bibliography 1893--1993},  Unpublished report ISRI-93-13, 17 June 1993, 19~pp.  

This bibliography is now superseeded by Refs.~\cite{GSPON2005D,GSPON2005E}.\\


~~\\ {\bf Additional references:} 

~~\\

\bibitem{R60} GSPONER A. and HURNI J.-P. (2001). \emph{Comment on Formulating and Generalizing Dirac's, Proca's, and Maxwell's Equations with Biquaternions or Clifford Numbers}, Found. Phys. Lett. {\bf 14}, 77--85. e-print arXiv:math-ph/0201049. 
 
\bibitem{R61} GSPONER A. (2002). \emph{On the ``equivalence'' of the Maxwell and Dirac equations}, Int. J. Theor. Phys. {\bf 41}, 689--694. e-print arXiv:math-ph/0201053.

\bibitem{R62} GSPONER A. (2002). \emph{Explicit closed-form parametrization of $SU(3)$ and $SU(4)$ in terms of complex quaternions and elementary functions}, Report ISRI-02-05  17~pp.  e-print arXiv:math-ph/0211056.

\bibitem{GSPON2005D} GSPONER A. and HURNI J.-P.  (2006). \emph{Quaternion in mathematical physics (1): Alphabetical bibliography}, Report ISRI-05-04 100~pp.  e-print arXiv:math-ph/0510059.

\bibitem{GSPON2005E} GSPONER A. and HURNI J.-P. (2006). \emph{Quaternion in mathematical physics (2): Analytical bibliography}, Report ISRI-05-05 113~pp.  e-print arXiv:math-ph/0511092.

\bibitem{GSPON2005F} GSPONER A. and HURNI J.-P. \emph{Quaternion in mathematical physics (3): Notations and terminology}, Report ISRI-05-06.

\bibitem{GSPON2005G} GSPONER A. and HURNI J.-P. \emph{Quaternion in mathematical physics (4): Formulas and methods}, Report ISRI-05-07.


\end{enumerate}
\end{document}